\documentclass[prd,aps,showpacs,nofootinbib,amsmath,amssymb]{revtex4}
\usepackage{amssymb,epsfig}
\usepackage[usenames,dvipsnames]{color}
\usepackage[bookmarks]{hyperref}
\usepackage[usenames,dvipsnames]{color}

\newtheorem{lemma}{Lemma}

\newtheorem{theorem}{Theorem}

\newcommand{\proof}{\noindent {\bf Proof. }}

\newcommand{\qed}{\hfill $\fbox{\hspace{0.3mm}}$ \vspace{.3cm}} %End of proof.

\newcommand{\Real}{\mathbb{R}}

\begin{document}
%%%%%%%%%%%%%%%%%%%%%%%%%%%%%%%

\title{Radial accretion flows on static spherically symmetric black holes}

\author{Eliana Chaverra$^{1,2}$ and Olivier Sarbach$^{1,2,3}$}
\affiliation{$^1$Instituto de F\'\i sica y Matem\'aticas,
Universidad Michoacana de San Nicol\'as de Hidalgo,\\
Edificio C-3, Ciudad Universitaria, 58040 Morelia, Michoac\'an, M\'exico,\\
$^2$Gravitational Physics, Faculty of Physics, University of Vienna, Boltzmanngasse 5, 1090 Vienna, Austria,\\
$^3$Perimeter Institute for Theoretical Physics, 31 Caroline St., Waterloo, ON, N2L 2Y5, Canada.}

\begin{abstract}
We analyze the steady radial accretion of matter into a nonrotating black hole. Neglecting the self-gravity of the accreting matter, we consider a rather general class of static, spherically symmetric and asymptotically flat background spacetimes with a regular horizon. In addition to the Schwarzschild metric, this class contains certain deformation of it which could arise in alternative gravity theories or from solutions of the classical Einstein equations in the presence of external matter fields. Modeling the ambient matter surrounding the black hole by a relativistic perfect fluid, we reformulate the accretion problem as a dynamical system, and under rather general assumptions on the fluid equation of state, we determine the local and global qualitative behavior of its phase flow. Based on our analysis and generalizing previous work by Michel, we prove that for any given positive particle density number at infinity, there exists a unique radial, steady-state accretion flow which is regular at the horizon. We determine the physical parameters of the flow, including its accretion and compression rates, and discuss their dependency on the background metric.
\end{abstract}

\date{\today}

% 04.20.-q: Classical GR
% 04.20.Ex: Initial value problem, existence and uniqueness of solutions
% 04.25.-g: Approximation methods; equations of motion
% 04.25.D-: Numerical relativity
% 04.25.Nx: Post-Newtonian approximation; perturbation theory; related approximations
% 04.40.-b: Self-gravitating systems, continuous media and classical fields in curved spacetime
% 04.70.-s: Physics of black holes
% 97.60.Lf: Astronomy: Late stage of star evolution: black holes
% 98.62.-g: Characteristics and properties of external galaxies and extragalactic objects
% 98.62.Mw: Infall, accretion, and accretion disks

\maketitle

%%%%%%%%%%%%%%%%%%%%%%%%%%%%%%%%%%%%%%%%%%%%%
\section{Introduction}
%%%%%%%%%%%%%%%%%%%%%%%%%%%%%%%%%%%%%%%%%%%%%

The accretion of gas into a black hole is an interesting physical process which has been studied extensively in the literature. As the gas falls into the black hole, effects due to compression or friction give rise to the emission of electromagnetic radiation whose spectrum can be observed and compared with theoretical models~\cite{Shapiro-Book}. A distinctive feature of black hole accretion (as opposed to accretion into a star) is the presence of an event horizon acting as a one-way membrane through which the accreting gas disappears. This has several important implications. First, the horizon provides a natural inner boundary condition for the equations of motion describing the gas, and thus complications or uncertainties related to finding the correct boundary conditions at the star's surface are avoided. Next, radiation emitted in the vicinity of the black hole undergo a strong gravitational lensing effect, resulting in an image showing the black hole's shadow surrounded by a sharp light ring. In fact, millimeter-wave very-long baseline interferometric arrays such as the Event Horizon Telescope~\cite{EHT} are already able to resolve the region around Sagittarius A$^*$, the supermassive black hole lying in the center of our galaxy, to scales smaller than its gravitational radius~\cite{sDetal08}, and thus may lead to new tests of general relativity in its strong field regime, see for example Ref.~\cite{aBtJaLdP14}.

In this article, we consider the simple case in which the black hole is nonrotating and the flow is radial and steady-state. The problem of spherical accretion of a fluid onto a gravity center has first been addressed by Bondi~\cite{hB52} in the Newtonian limit. Generalizations to relativistic spherical steady-states on the Schwarzschild geometry and applications to the accretion of matter from the interstellar medium into a nonrotating black hole have been given by Michel~\cite{fM72} and further discussed in Ref.~\cite{Shapiro-Book}. Given a fluid equation of state and the temperature of the gas at infinity, regularity of the flow at the black hole horizon singles out a unique solution which turns out to be transonic, the sonic sphere describing the transition of the flow's radial velocity measured by static observers from sub- to supersonic lying outside the horizon. The stability of the Michel flow in the subsonic region outside the sonic sphere has been established by Moncrief~\cite{vM80}. For more recent work on radial flows, see Ref.~\cite{dAsBtD14a} for a perturbative study, Refs.~\cite{jKeM13,pMeM13,pMeMjK13} for the study of radial accretion onto cosmological black holes, and Ref.~\cite{fGfL11} for the accretion of dark matter into a supermassive black hole. For the self-gravitating case, see Refs.~\cite{eM99,fLmGfG14}.

The purpose of this work is to provide rigorous results regarding the existence and uniqueness of the Michel flow, and to generalize it to equations of state more general than the simple polytrope and to background metrics more general than the Schwarzschild one. In fact, we consider a rather general class of static, spherically symmetric and asymptotically flat spacetimes which have a regular Killing horizon and whose metric coefficients obey certain monotonicity conditions. As a consequence, our results are also applicable to nonrotating black holes predicted by alternative theories of gravity or to distorted black holes in Einstein gravity in the presence of external fields, like dark matter for example.

We study the accretion problem by casting it into a Hamiltonian dynamical system on the phase space $(r,n)$ parametrized by the areal radius coordinate $r$ and the particle density $n$ of the fluid. Although the Hamiltonian describing this system is fictitious and does not have the standard ``kinetic plus potential energy" form, nevertheless the system can be analyzed using standard tools from the theory of dynamical systems. Based on our assumptions on the fluid equation of state and the metric fields we give a complete description for the qualitative behavior of the phase flow. In particular, we show that there exists a unique hyperbolic critical point of this system which corresponds to a saddle point of the Hamiltonian. The global behavior of the stable and unstable one-dimensional manifolds through this critical point are discussed, and we prove that the unstable manifold describes the unique flow which extends from the horizon to infinity and has a finite, positive particle density at infinity. This solution describes the Michel flow and the critical point through which it passes corresponds to the sonic sphere. Conversely, we prove that for a given positive particle density at infinity there exists a unique critical point of the system whose unstable manifold matches it. Finally, assuming a polytropic equation of state with adiabatic index $\gamma$ lying in the range $1 < \gamma \leq 5/3$ and focusing on the case where the fluid's sound speed at infinity $v_\infty$ is much smaller than the speed of light $c$, we compute the accretion rate as a function of the particle density at infinity and the compression rates of the fluid at the horizon and the sonic sphere. For a gas with $1 < \gamma < 5/3$ we find that the compression rate at the sonic sphere does not depend on the details of the background metric to leading order in $v_\infty/c$, though a nontrivial dependency appears to next-to-leading order. However, the compression rate at the event horizon depends quadratically on the ratio $\sigma_1 = r_s/r_H$, where $r_s = 2GM/c^2$ is the Schwarzschild radius of the black hole and $r_H$ its event horizon radius ($\sigma_1 = 1$ in the Schwarzschild case). For the case where the adiabatic index is $\gamma = 5/3$ both the compression rates at the sonic sphere and the horizon depend on the details of the metric to leading order in $v_\infty/c$. The accretion rate, on the other hand, is independent of the details of the background metric and is given by the familiar Bondi formula which only depends on the total mass $M$ of the black hole, the particle density at infinity and $v_\infty/c$ to leading order in $v_\infty/c$.

The remaining part of this paper is structured as follows: in Sec.~\ref{Sec:Model} we first define the class of static, spherically symmetric black hole metrics we are considering in this work. Then, we review the derivation for the equations of motion describing a steady-state, radial perfect fluid flow, specify our assumptions on the fluid equation of state and enunciate our main results. In Sec.~\ref{Sec:PhaseSpace} we reformulate the fluid equations as a (fictitious) two-dimensional Hamiltonian dynamical system on the phase space $(r,n)$, where the Hamiltonian depends on the accretion rate $\mu$. We perform a detailed analysis of the phase flow, starting with the standard local analysis, and show that under our assumptions and for large enough $|\mu|$ there exists a unique hyperbolic critical point $(r_c,n_c)$, with associated stable and unstable local one-dimensional manifolds. Next, we analyze the global behavior of the phase flow and prove that the unstable manifold extends from the horizon to the asymptotic region, assigning to each value of $r$ a unique value for the particle density $n$. This solution describes the Michel flow, and the critical point it passes through corresponds, physically, to the sonic sphere. We also show in Sec.~\ref{Sec:PhaseSpace} that given a positive particle number $n_\infty  > 0$ at infinity, there exists a unique value for $|\mu|$ and a unique critical point $(r_c,n_c)$ such that the corresponding Michel flow solution matches the given value of $n_\infty$ as $r\to\infty$. We also make brief comments on other solutions which could be useful when describing stellar winds or subsonic flows accreted by relativistic stars. Next, in Sec.~\ref{Sec:Examples} we discuss specific examples, including the accretion of dust and of a stiff fluid, provide numerical plots for the phase flow of polytropic fluids on a Schwarzschild and other backgrounds, and analyze the parameters characterizing the accretion flow under the usual assumption that the speed of sound at infinity is much smaller than the speed of light. For the case of a polytropic equation of state, we compute the accretion rate and the compression rates of the fluid at the sonic sphere and the event horizon for small $v_\infty/c$, and analyze their dependency on the background metric. Conclusions are drawn in Sec.~\ref{Sec:Conclusions}. Some physical justification for our assumptions on the metric fields are discussed in Appendix~\ref{App:MetricConditions}, while the proofs of some technical results needed in our analysis can be found in Appendix~\ref{App:Proofs}.

%%%%%%%%%%%%%%%%%%%%%%%%%%%%%%%%%%%%%%%%%%%%
\section{Steady-state radial accretion flows into a black hole and main results}
\label{Sec:Model}
%%%%%%%%%%%%%%%%%%%%%%%%%%%%%%%%%%%%%%%%%%%%

In this section, we specify our assumptions on the background black hole metric and on the fluid equation of state, review the relevant equations describing a relativistic, steady, radial flow, and state our main results regarding the existence and main properties of such flows.

\subsection{Assumptions on the metric field}

The background metric on which we study the accretion in this work is modeled by a spherically symmetric metric of the form
\begin{equation}
ds^2 = -\alpha(r)^2 N(r) c^2 dt^2  + \frac{dr^2}{N(r)}
 + r^2\left( d\vartheta^2 + \sin^2\vartheta d\varphi^2 \right),
\label{Eq:metric}
\end{equation}
where $\alpha$ and $N$ are smooth functions of the areal radius $r$ which are independent of the time coordinate $t$ and the spherical coordinates $(\vartheta,\varphi)$. Accordingly, the spacetime described by the metric~(\ref{Eq:metric}) is invariant with respect to the flow defined by the Killing vector field ${\bf k} = \frac{1}{c}\frac{\partial}{\partial t}$ and with respect to rotations on the two-sphere. The geometric interpretation of the functions $\alpha$ and $N$ is the following: the Killing vector field ${\bf k}$ and the differential $dr$ of the areal radius give rise to the two scalar quantities $\sigma := -{\bf g}({\bf k},{\bf k}) = -g_{\mu\nu} k^\mu k^\nu$ and ${\bf g}(dr,dr) = g^{\mu\nu}(\nabla_\mu r)(\nabla_\nu r)$. In terms of these scalar fields, $N = {\bf g}(dr,dr)$ and $\alpha = \sqrt{\sigma/N}$. For the Schwarzschild metric, $\alpha = 1$ and $N(r) = \sigma(r) = 1 - r_H/r$, and the event horizon is located at $r = r_H$, where $r_H = 2GM/c^2$ with $M$ the black hole's mass and $G$ Newton's constant.

In this work, we consider a more general class of static, spherically symmetric black holes which is characterized by the following requirements on the smooth functions $\alpha, N: (0,\infty)\to \Real$:
\begin{enumerate}
\item[(M1)] {\bf Asymptotic flatness with positive total mass:}\\
As $r\to\infty$, $\alpha(r)\to 1$, $N(r)\to 1$, and $r^2\sigma'(r) \to 2m $ for some positive constant $m > 0$.
\item[(M2)] {\bf Regular Killing horizon:}\\
There exists a radius $r_H > 0$ such that $\sigma(r_H) = 0$, $\sigma'(r_H) > 0$ and $\alpha(r_H) > 0$.
\item[(M3)] {\bf Monotonicity conditions:}\\
The functions $\alpha$, $\sigma$ and $\sigma'$ are strictly positive on the interval $r > r_H$.
\item[(M4)] {\bf Proximity to Schwarzschild condition:}\\
The following inequality holds for all $r > r_H$:
\begin{equation}
-3 < \frac{(r^2\sigma')'}{r\sigma'} < \frac{m}{r}\frac{9}{8 + \frac{m}{r}}.
\label{Eq:M4}
\end{equation}
\end{enumerate}
Before we proceed, let us briefly discuss the meaning of these conditions. For more details, definition and examples we refer the reader to Appendix~\ref{App:MetricConditions}. Condition (M1) assures that the spacetime described by the metric~(\ref{Eq:metric}) is asymptotically flat in the sense that it approaches the Minkowski metric as $r\to\infty$, and that its total Komar mass is $M = mc^2/G > 0$. Next, condition (M2) implies that the metric possesses a Killing horizon at $r = r_H$, where the Killing vector field ${\bf k}$ is null. The associated surface gravity is
\begin{equation}
\kappa = \frac{c^2}{2}\frac{\sigma'(r_H)}{\alpha(r_H)} > 0,
\end{equation}
implying that the horizon is non-degenerated. Next, condition (M3) can be motivated by requiring the effective stress-energy tensor obtained by plugging the metric in Eq.~(\ref{Eq:metric}) into Einstein's equations, to satisfy the weak and the strong energy conditions. Finally, for the meaning of the last condition we note that for the Schwarzschild metric, $\sigma(r) = 1 - r_H/r$ such that $r^2\sigma' = r_H$ is constant, implying that the inequality in Eq.~(\ref{Eq:M4}) is trivially satisfied in this case. Therefore, the condition (M4) can be interpreted as a restriction on the size of deformation of the function $\sigma$ relative to its Schwarzschild value. An explicit example for such a deformation is given in Appendix~\ref{App:MetricConditions}.

\subsection{Radial perfect fluid flows}

After having specified our assumptions on the background geometry, we now turn to the properties of the radial flow considered in this work. As described in the introduction, we model the flow by a relativistic perfect fluid, thereby neglecting effects related to viscosity or heat transport, and further assume that the fluid's energy density is sufficiently small so that its self-gravity can be neglected. We also assume that the fluid is minimally coupled to gravity, so that it satisfies the standard relativistic Euler equations.\footnote{See, for instance, Ref.~\cite{vF09} for modified gravity theories with non-minimally coupled perfect fluids, in which case a term involving the gradient of the Ricci scalar would appear on the right-hand side of Eq.~(\ref{Eq:consEner}).} The fluid is described by a four-velocity field ${\bf u} = u^\mu\partial_\mu$, normalized such that $u^\mu u_\mu = -c^2$, and the particle density $n$, energy density $\varepsilon$, pressure $p$, temperature $T$, and entropy per particle $s$ measured by comoving observers. For the following, the enthalpy per particle, defined as $h:= (\varepsilon + p)/n$, will play an important role. We require local thermodynamic equilibrium and the existence of an equation of state $h = h(p, s)$ satisfying the first law of thermodynamics
\begin{equation}
dh = T ds + \frac{dp}{n}.
\label{Eq:FirstLaw}
\end{equation}

The dynamics of the fluid flow is governed by the equations
\begin{eqnarray}
\nabla_\mu J^\mu &= &0,
\label{Eq:consCorri}\\
\nabla_\mu T^{\mu\nu} &=& 0.
\label{Eq:consEner}
\end{eqnarray}
wherein $J^\mu = n u^\mu$ is the particle current density and $T^{\mu\nu} = n h u^\mu u^\nu + p g^{\mu\nu}$ is the stress-energy tensor, and $\nabla$ refers to the covariant derivative with respect to the spacetime metric ${\bf g}$. Eq.~(\ref{Eq:consCorri}), the component of Eq.~(\ref{Eq:consEner}) along ${\bf u}$ and the first law, Eq.~(\ref{Eq:FirstLaw}), yield
\begin{equation}
T u^\mu\nabla_\mu s = 0,
\label{Eq:consHeat}
\end{equation}
implying that there is no heat transfer between the different fluid elements.

For the particular case of a radial flow, the four-velocity ${\bf u}$ has vanishing angular components, so that ${\bf u} = u^t\partial_t + u^r\partial_r = u^0{\bf k} + u^r\partial_r$, with $u^0 = c u^t$. The four-velocity's radial component $u^r = dr({\bf u})$ is negative in the accreting case. The radial component $v$ of the fluid's three-velocity measured by static observers is defined by
\begin{equation}
{\bf u} = \frac{1}{\sqrt{1-\beta^2}} \left( c {\bf e}_0 + v {\bf e}_1 \right),
\qquad \beta := \frac{v}{c},
\end{equation}
where ${\bf e}_0 := {\bf k}/\sqrt{\sigma}$ is the static observer's four-velocity and ${\bf e}_1 := \sqrt{N}\partial_r$ is a unit radial vector field orthogonal to it. Consequently, the two velocities $u^r$ and $v$ are related to each other via
\begin{equation}
u^r = \sqrt{N(r)}\frac{v}{\sqrt{1 - \beta^2}}.
\label{Eq:uder}
\end{equation}
While $u^r$ is well-defined everywhere, the velocity $v$ is only defined \emph{outside} the horizon, where $N(r)$ is positive, since there are no static observers inside the black hole. A third velocity which will play a very important role in the description of the fluid flow is the fluid's local speed of sound $v_s$, defined as
\begin{equation}
\frac{v_s^2}{c^2} := \left. \frac{\partial p}{\partial\epsilon} \right|_s
 = \left. \frac{\partial\log h}{\partial\log n} \right|_s. 
\end{equation}

After these remarks, we derive the equations of motion describing a steady-state radial flow on the spherically symmetric static black hole background described by Eq.~(\ref{Eq:metric}). Since the fluid is steady-state and radial, the quantities $u^r$, $n$, $h$, $v_s$, $p$, $s$ describing the fluid depend on the areal radius coordinate $r$ only. Eq.~(\ref{Eq:consHeat}) immediately implies that $s = const$; hence each fluid element has the same entropy and it is sufficient to consider an equation of state of the form $h(n)$. Next, the particle conservation law, Eq.~(\ref{Eq:consCorri}), yields
\begin{equation}
\frac{d}{dr} \left( \alpha r^2 n u^r \right) = 0,
\end{equation}
from which it follows that the particle flux through a sphere of constant areal radius $r$,
\begin{equation}
j_n := 4\pi \alpha r^2 n u^r = const.,
\label{Eq:jn}
\end{equation}
is independent of $r$. (Notice that $j_n < 0$ is negative for accretion.) Another conservation law is obtained by contracting Eq.~(\ref{Eq:consEner}) with $k_\nu$ and using the Killing equation $\nabla_\mu k_\nu + \nabla_\nu k_\mu = 0$. This yields
\begin{equation}
\frac{d}{dr} \left( \frac{\alpha r^2}{c} T^r{}_t \right) = 0,
\end{equation}
which implies the conservation of the energy flux $j_\varepsilon := -4\pi\alpha r^2 T^r{}_t/c$ through the spheres of constant $r$. Using $T^r{}_t = n h u^r u_t$ and the normalization condition of ${\bf u}$, we obtain
\begin{equation}
j_\varepsilon = 4\pi\alpha^2 r^2 n h u^r\sqrt{N + \left( \frac{u^r}{c} \right)^2} = const.
\label{Eq:je}
\end{equation}
Summarizing,\footnote{Note that it is sufficient to consider Eq.~(\ref{Eq:consCorri}) and the components of Eq.~(\ref{Eq:consEner}) along ${\bf u}$ and ${\bf k}$, since the angular components of Eq.~(\ref{Eq:consEner}) are trivial as a consequence of spherical symmetry.} the equations of motion describing a radial, steady-state perfect fluid flow on a static, spherically symmetric background consist of the two conservation laws~(\ref{Eq:jn},\ref{Eq:je}), together with an equation of state $h = h(n)$.

Taking the quotient of Eqs.~(\ref{Eq:jn}) and (\ref{Eq:je}), and eliminating $u^r$ from Eq.~(\ref{Eq:jn}) leads to the following implicit problem:
\begin{equation}
F(r,n) := h(n)^2 \left[ \sigma(r) + \frac{\mu^2}{r^4 n^2} \right]
 = \left(\frac{j_\varepsilon}{j_n}\right)^2 = const.,\qquad
\mu := \frac{j_n}{4\pi c},
\label{Eq:Fundamental}
\end{equation}
for the particle density $n$, where we recall that the function $\sigma(r) = \alpha(r)^2 N(r)$ is given by the metric fields and is subject to the assumptions (M1)--(M4). Therefore, the problem of determining the flow is reduced to the problem of finding an appropriate level curve of the function $F$ which associates to each value of $r$ a unique value of the particle density $n(r)$. Once $n(r)$ has been determined, the radial velocity $u^r$ is obtained from Eq.~(\ref{Eq:jn}),
\begin{equation}
\frac{u^r(r)}{c} = \frac{\mu}{\alpha(r) r^2 n(r)},
\label{Eq:urVelocity}
\end{equation}
from which
\begin{equation}
\frac{v(r)}{c} = \frac{\mu}{\sqrt{\mu^2 + r^4\sigma(r) n(r)^2}}.
\label{Eq:RadialVelocity}
\end{equation}

\subsection{Assumptions on the fluid equation of state and main results}

So far, the function $h(n)$, describing the fluid equation of state, has been left unspecified. We now formulate our conditions on $h$ and state our main results concerning the solutions of Eq.~(\ref{Eq:Fundamental}). Their proofs are given in the following section.

We assume $h: (0,\infty)\to (0,\infty)$ to be a smooth function satisfying the following properties:
\begin{enumerate}
\item[(F1)] {\bf Positive rest energy:}\\
As $n\to 0$, $h(n)\to e_0$ converges to a positive number $e_0 > 0$.
\item [(F2)] {\bf Positive and subluminal sound speed:}\\
$0 < v_s(n) < c$ for all  $n > 0$.
\item [(F3)] {\bf Technical restriction on the derivative of $v_s(n)$:}\\
$0\leq W(n):=\frac{\partial\log v_s}{\partial\log n}\leq \frac{1}{3}$ for all $n > 0$.
\end{enumerate}
While the first two conditions are physically well-motivated, the meaning of the third condition is not immediately obvious. As we will see in Secs.~\ref{Sec:PhaseSpace} and~\ref{Sec:Examples} it is required to ensure the existence of a unique critical point in the phase flow. For the particular case of a polytropic equation of state,
\begin{equation}
p = k n^\gamma,
\end{equation}
with $k$ a positive constant and $\gamma$ the adiabatic index, integration of the first law, Eq.~(\ref{Eq:FirstLaw}), with $ds = 0$ gives
\begin{equation}
h(n) = e_0 + \frac{\gamma k}{\gamma-1} n^{\gamma-1}
\end{equation}
with $e_0$ the rest energy of the particle. It is easily verified that the conditions (F1)--(F3) are satisfied in this case, provided $e_0 > 0$ and $1 < \gamma \leq 5/3$. However, more general equations of state with variable adiabatic index $\gamma$ are also contained in our assumptions~(F1)--(F3).

Before we state our main results concerning the solution of Eq.~(\ref{Eq:Fundamental}), let us make the following observation: assume there exists a smooth solution $n(r)$ of Eq.~(\ref{Eq:Fundamental}) which is well-defined on the interval $[r_H,\infty)$ including the event horizon at $r = r_H$ and which converges to a finite, positive value $n_\infty$ at infinity. By differentiating $F(r,n(r)) = const.$ with respect to $r$ we obtain
\begin{equation}
\frac{\partial F}{\partial r}(r,n(r)) + \frac{\partial F}{\partial n}(r,n(r)) n'(r) = 0.
\label{Eq:nprime}
\end{equation}
The implicit function theorem guarantees the local existence and uniqueness of $n(r)$ as long as
$$
\frac{\partial F}{\partial n}(r,n) 
 = \frac{2h(n)^2}{n}\left[ \frac{v_s^2}{c^2}\sigma(r) 
 - \left( 1 - \frac{v_s^2}{c^2} \right)\frac{\mu^2}{r^4 n^2} \right]
$$
is different from zero. According to our assumptions, the right-hand side is negative at the horizon (where $\sigma = 0$) and positive for large $r$ (where $\sigma$ is close to one). Therefore, while the hypotheses of the implicit function theorem are satisfied close to the horizon and in the asymptotic region, they are violated at some intermediate point $r = r_c$, where $\partial F/\partial n(r_c,n_c) = 0$, $n_c := n(r_c)$. In view of Eq.~(\ref{Eq:nprime}) the only possibility for $n(r)$ to have a well-defined first derivative at $r = r_c$ is that the partial derivative of $F$ with respect to $r$ also vanishes at $(r_c,n_c)$. In other words, \emph{a necessary condition for obtaining a differentiable level curve $n(r)$ of Eq.~(\ref{Eq:Fundamental}) which extends from the horizon to infinity and has a finite, positive limit $n_\infty$ as $r\to\infty$ is that this curve passes through a critical point of $F$.}

Our first main result is concerned with the existence and uniqueness of the critical point $(r_c,n_c)$ of $F$ and the structure of the level curves passing through it.

\begin{theorem}[Critical point and level curves passing through it]
\label{Thm:Main}
Under our assumptions (M1)--(M4) on the background metric and (F1)--(F3) on the fluid equation of state, the function $F: [r_H,\infty)\times (0,\infty)\to\Real, (r,n)\mapsto F(r,n)$ defined by Eq.~(\ref{Eq:Fundamental}) possesses a critical point $(r_c,n_c)$ for each large enough value of $|\mu|$. The critical point (when it exists) is unique and describes a saddle point of $F$ at which two level curves of $F$ cross. These level curves can be parametrized by $(r,n_+(r))$ and $(r,n_-(r))$, $r > r_H$, respectively, with the functions $n_\pm$ satisfying the following properties:
\begin{enumerate}
\item[(a)] $n_+ : (r_H,\infty)\to\Real$ is the unique differentiable function satisfying $n_+(r_c) = n_c$ and $F(r,n_+(r)) = F(r_c,n_c)$ for all $r > r_H$ which has regular limits as $r\to r_H$ and $r\to \infty$. Moreover, $n_\infty := \lim_{r\to\infty} n_+(r) > 0$.
\item[(b)] $n_- : (r_H,\infty)\to\Real$ is the unique differentiable function satisfying $n_-(r_c) = n_c$ and $F(r,n_-(r)) = F(r_c,n_c)$ for all $r > r_H$ such that $\lim_{r\to\infty} n_-(r) = 0$.
\end{enumerate}
Both functions $n_\pm$ are monotonously decreasing.
\end{theorem}

The function $n_+$ describes the Michel flow, a radial steady-state accretion flow which extends from the event horizon of the black hole to the asymptotic region and which has vanishing radial velocity and finite, positive particle density $n_\infty$ at infinity. An example plot showing the contours of the function $F$ and the level curves $(r,n_\pm(r))$ for the particular case of a polytrope on a Schwarzschild background are shown in Fig.~\ref{Fig:PS} below. As will become clear from the proof in the next section, the flow is subsonic for $r > r_c$ and supersonic for $r < r_c$, the critical radius $r_c$ representing the sonic sphere. In contrast to this, as we will see, the function $n_-$ diverges as $r\to r_H$ and thus it is not well-defined at the event horizon. The corresponding flow is subsonic for $r_H < r < r_c$, supersonic for $r > r_c$, and at infinity the particle density vanishes and the radial speed is positive. When the black hole is replaced by a compact, spherical star this solution plays an important role in the description of relativistic stellar winds~\cite{eP65}.

The functions $n_\pm$ in Theorem~\ref{Thm:Main} parametrizing the level set $F(r,n) = F(r_c,n_c)$ corresponding to the critical point $(r_c,n_c)$ depend on the accretion rate parameter $\mu$, see Eq.~(\ref{Eq:Fundamental}). Instead of parametrizing the Michel flow by $\mu$ one can also parametrize it in terms of its particle density at infinity $n_\infty$. This is the statement of our second main result:

\begin{theorem}
\label{Thm:MainBis}
Let the conditions (M1)--(M4) on the background metric and (F1)--(F3) on the fluid equation of state be satisfied. Then, given any $n_\infty > 0$, there exists a unique value of $|\mu|$ such that the function $F$ has a critical point with the property that the function $n_+$ defined in Theorem~\ref{Thm:Main} satisfies $\lim_{r\to\infty} n_+(r) = n_\infty$.
\end{theorem}

The value of $n_\infty$ at infinity can be determined by the particle density of the ambient matter far from the black hole. For the particular scenario of a black hole accreting gas from the interstellar medium, it is common to fit the flow's temperature $T_\infty$ to the temperature of the interstellar medium, which in turn is related to the sound speed at infinity $v_\infty$ via the equation of state. For nonrelativistic baryons, for example, we have~\cite{Shapiro-Book}
$$
n k_B T = p = k n^{5/3},
$$
which, taking into account the proton rest mass $m_p = 938MeV/c^2$, leads to
$$
v_\infty = \left( \frac{5}{3}\frac{k_B T_\infty}{m_p} \right)^{1/2} 
 \simeq 11.7\left( \frac{T_\infty}{10^4 K} \right)^{1/2} \frac{km}{s},
$$
which is considerably smaller than the speed of light for typical temperatures of $T_\infty \simeq 10^4 K$.

When $v_\infty/c\ll 1$ we expect that the sonic sphere is located far from the event horizon, $r_c \gg r_H$. Assuming a polytropic equation of state with adiabatic index $1 < \gamma \leq 5/3$, we confirm this expectation in Sec.~\ref{Sec:Examples}, where we compute the leading order contributions in $v_\infty/c$ to the parameters characterizing the flow. For the particular case $\gamma = 5/3$ mentioned above we find the following expressions for the compression rates to leading order in $v_\infty/c$:
\begin{eqnarray}
\frac{n_c}{n_\infty} &\simeq& \left( \frac{2}{3l}\right)^{3/2} 
\left( \frac{v_\infty}{c} \right)^{-3/2}
 \simeq 2.224\times10^6 l^{-3/2}\left(\frac{T_\infty}{10^4K}   \right)^{-3/4},
\label{Eq:CRnc}\\
\frac{n_H}{n_\infty} &\simeq& \left(\frac{2}{3}\right)^{3/2}z_{H0}
\left( \frac{v_\infty}{c} \right)^{-3}
 \simeq 1.334\times10^{12}\left( \frac{z_{H0}}{0.14677} \right)
 \left(\frac{T_\infty}{10^4K}   \right)^{-3/2},
\label{Eq:CRnH}
\end{eqnarray}
where the dimensionless quantities $\ell$ and $z_{H0}$ are defined in Eqs.~(\ref{Eq:l}) and~(\ref{Eq:zH0Solution}) below. They depend on the parameters $\sigma_1 = 2m/r_H$ and $\sigma_2$ in the asymptotic expansion
$$
\sigma(r) = 1 - \sigma_1\frac{r_H}{r} - \sigma_2\frac{r_H^2}{r^2}
 + {\cal O}\left( \frac{r_H^3}{r^3} \right)
$$
of the metric function $\sigma$. For a Schwarzschild black hole, $l = 1$ and $z_{H0} \simeq 0.14677$, and the compression rates in Eqs.~(\ref{Eq:CRnc},\ref{Eq:CRnH}) reduce to the known results in the literature~\cite{fM72,Shapiro-Book}; however these values may be significantly different for distorted black holes. The accretion rate is given to leading order in $v_\infty/c$ by:
\begin{equation}
j_\varepsilon \simeq \pi m_p c^3\left( \frac{G M}{c^2} \right)^2
\left( \frac{v_\infty}{c} \right)^{-3} n_\infty
 \simeq 5.158\times 10^{24}\left(\frac{M}{M_\odot}\right)^2
 \left(\frac{T_\infty}{10^4K}   \right)^{-3/2}\left( \frac{n_\infty}{1 cm^{-3} } \right) \frac{J}{s},
\end{equation}
and only depends on the total mass $M$ of the black hole, not on the details of the metric.

In the next section, we prove Theorems~\ref{Thm:Main} and \ref{Thm:MainBis} by reformulating the implicit problem in Eq.~(\ref{Eq:Fundamental}) as a (fictitious) Hamiltonian dynamical system on the phase space $(r,n)$. The proof generalizes our previous work~\cite{eCoS12} (see also Refs.~\cite{eM99,Eliana-Master-thesis}) for the particular case of a polytropic fluid accreting on a Schwarzschild black hole background. A reader who is not interested in the proof may skip this section and proceed to Sec.~\ref{Sec:Examples}, where we discuss specific examples involving different equations of states and background black holes.

%%%%%%%%%%%%%%%%%%%%%%%%%%%%%%%%%%%%%%%%%%%%
\section{Phase space analysis}
\label{Sec:PhaseSpace}
%%%%%%%%%%%%%%%%%%%%%%%%%%%%%%%%%%%%%%%%%%%%

The main objective of this section is to prove Theorems~\ref{Thm:Main} and \ref{Thm:MainBis}. In a first step, we rewrite the fundamental equation~(\ref{Eq:Fundamental}) in terms of dimensionless quantities and state some preliminary results regarding the behavior of the metric fields and the specific enthalpy function in Sec.~\ref{SubSec:Preliminary}. Next, in Sec.~\ref{SubSec:DS} we reformulate the problem in terms of a dynamical system on the phase space $(r,n)$, where the vector field describing the dynamics is the Hamiltonian vector field associated with the function $F$ defined in Eq.~(\ref{Eq:Fundamental}). By construction, $F$ is constant along the phase trajectories and thus the latter describe its level curves. The main advantage of casting the problem into a dynamical system relies on the fact that the behavior near the critical point of $F$ (where the hypotheses of the implicit function theorem break down) can be analyzed using standard tools from the theory of dynamical systems~\cite{Hartman-Book,Perko-Book}. Using local analysis, we show in Sec.~\ref{SubSec:DS} that for large enough values of $|\mu|$ the Hamiltonian flow possesses a unique hyperbolic critical point with associated stable and unstable local one-dimensional manifolds, which are defined as the set of points lying in a vicinity of the critical point which converge to it in the future and past, respectively, along the phase flow. In Sec.~\ref{SubSec:DSGlobal} we examine the extensions of these manifolds and prove that the unstable branch extends all the way from the horizon to the asymptotic region, and thus it describes the physical relevant solution. Our analysis in Sec.~\ref{SubSec:DSGlobal} also reveals the global structure of the phase space, and in particular proves Theorem~\ref{Thm:Main}. Finally, the statement of Theorem~\ref{Thm:MainBis} is proved in Sec.~\ref{SubSec:Matching}. The proofs of some technical Lemmata needed for our analysis can be found in Appendix~\ref{App:Proofs}.

%%%%%%%%%%%%%%%%%%%%%%%%%%%%%%%%%%%%%%%%%%%%
\subsection{Dimensionless quantities and preliminary results}
\label{SubSec:Preliminary}

We introduce the following dimensionless quantities:
\begin{eqnarray}
&& x := \frac{r}{r_H},\quad
     u := \alpha\frac{u^r}{c},\quad
     z := \frac{n}{n_0},\quad
     \bar{\mu} := \frac{j_n}{4\pi c r_H^2 n_0}\\
&& f(z) := \frac{h(n)}{e_0},\quad
      \nu^2(z) := \frac{\partial \log f(z)}{\partial \log z},\quad
      w(z) := \frac{\partial \log\nu(z)}{\partial \log z}
\end{eqnarray}
Here, $n_0$ denotes a typical density (which could be defined as the particle density at infinity, for example), $\bar{\mu}$ is the dimensionless accretion rate, $\nu$ is the local speed of sound divided by $c$, and $f$ denotes the specific enthalpy. Note that in the accreting case, both $\bar{\mu}$ and $u$ are negative. For notational simplicity we shall drop the bar on $\bar{\mu}$ in the following. In terms of these quantities, the fundamental equation, Eq.~(\ref{Eq:Fundamental}), can be rewritten as
\begin{equation}
F_\mu(x,z) = \left(\frac{j_\varepsilon}{j_n e_0}\right)^2 = const.,
\label{Eq:FundamentalBis}
\end{equation}
where $F_\mu: \Omega \to \Real$ is the function on $\Omega:=(0,\infty)\times (0,\infty)$ defined by
\begin{equation}
F_\mu(x,z) := f(z)^2 \left[ \sigma(x) + \frac{\mu^2}{x^4 z^2} \right],\qquad
(x,z)\in\Omega.
\label{Eq:Fmu}
\end{equation}
For the analysis of the Hamiltonian flow associated with the function $F_\mu$ we will need two technical lemmata. The first one is related to the behavior of the metric field $\sigma$.

\begin{lemma}
\label{Lem:Technical1}
The function
$$
a(x) := \frac{\sigma(x)}{x\sigma'(x)},\qquad x\geq 1,
$$
is a monotonously increasing function satisfying $a(1) = 0$, $\lim_{x\to\infty} a(x)/x = r_H/(2m)$ and the inequalities
\begin{equation}
0 < a(x) < \frac{r_H}{2m}x,\qquad
\frac{(4a+1)^2}{16a+1} < x a' < 1 + 4a,
\end{equation}
for all $x > 1$.
\end{lemma}

\proof See Appendix~\ref{App:Proofs}.
\qed

The second lemma, which is a consequence of the assumptions (F1)--(F3) on the fluid equation of state, gives information about the asymptotic properties of the specific enthalpy $f$ and the dimensionless sound speed $\nu$ as $z\to 0$ and $z\to \infty$.

\begin{lemma}
\label{Lem:Technical2}
\begin{enumerate}
\item[(a)] Given $z_1 > 0$ there exists $\delta > 0$ such that the following inequalities hold for all $0 \leq z \leq z_1$:
\begin{equation}
\nu(z) \geq \delta z^{1/3},\qquad
f(z) \geq \exp\left[ \frac{3}{2}\delta^2 z^{2/3} \right].
\end{equation}
\item[(b)] $\nu$ is a strictly monotonously increasing function satisfying
$$
\lim\limits_{z\to 0}\nu(z) = 0,\qquad
\lim\limits_{z\to\infty}\nu(z) = \nu_1
$$
for some constant $0 < \nu_1 \leq 1$.
\item[(c)] Given $z_0 > 0$ there exists a constant $q > 0$ such that for all $z \geq z_0$,
\begin{equation}
f(z) \geq f(z_0)\left( \frac{z}{z_0} \right)^q.
\end{equation}
In particular, $\lim\limits_{z\to\infty} f(z) = \infty$.
\end{enumerate}
\end{lemma}

\proof See Appendix~\ref{App:Proofs}.
\qed

%%%%%%%%%%%%%%%%%%%%%%%%%%%%%%%%%%%%%%%%%%%%
\subsection{Definition of the dynamical system and critical points}
\label{SubSec:DS}

The Hamiltonian vector field with respect to the function $F_\mu$ is defined by
\begin{equation}
X_F(x,z) := (\nabla F_\mu(x,z))^\perp =
\left( \begin{array}{r}
 \frac{\partial F_\mu(x,z) }{\partial z}  \\   -\frac{\partial F_\mu(x,z)}{\partial x} 
\end{array} \right),\qquad (x,z)\in \Omega.
\label{Eq:HamVec}
\end{equation}
The associated Hamiltonian flow is determined by the integral curves of $X_F$, obtained by solving the ordinary differential system
\begin{equation}
\frac{d}{d\lambda}
\left( \begin{array}{r}
 x  \\  z
\end{array} \right)
= X_F(x,z)
 = f(z)^2\left. \left( \begin{array}{c}
 \frac{2}{z}\left[ \nu(z)^2\left( \sigma(x)+ u^2 \right) - u^2 \right] \\
 -\sigma'(x) + \frac{4 u^2}{x}
\end{array} \right) \right|_{u = \frac{\mu}{x^2 z}},
\label{Eq:DynSys}
\end{equation}
with $\lambda$ a fictitious time parameter. By construction, these curves leave the level sets of $F_\mu$ invariant, so that the problem of finding implicit solutions of the equation $F_\mu(x,z) = const.$ can be understood by determining the qualitative properties of the phase flow.

The critical points $(x_c,z_c)$ of the system are determined by the points at which $X_F$ vanishes, that is
\begin{eqnarray}
\nu(z)^2\left( \sigma(x)+ u^2 \right) - u^2  &=& 0,
\label{Eq:CritCond1}\\
\sigma'(x) - \frac{4 u^2}{x} &=& 0,
\label{Eq:CritCond2}
\end{eqnarray}
where $u^2 = \mu^2/(x^4 z^2)$. Eliminating $z$, solving this system is equivalent to finding the zeros of the auxiliary function
\begin{equation}
\mathcal{F}_\mu(x):= \nu(z(x))^2\left[ 1 + 4a(x) \right] - 1,\qquad
z(x) := \frac{2|\mu|}{\sqrt{x^5\sigma'(x)}},\qquad
x\geq 1,
\label{Eq:AuxiliaryFct}
\end{equation}
where $a(x) = \sigma(x)/(x\sigma'(x))$ is the function defined in Lemma~\ref{Lem:Technical1}. We first show the existence of a zero by analyzing the limits $x\to 1$ and $x\to \infty$. For $x\to 1$,
\begin{displaymath}
\mathcal{F}_\mu(1) = \nu\left( 2|\mu|/\sqrt{\sigma'(1)} \right)^2 - 1 < 0,
\end{displaymath}
where we have used the assumptions (M2) and (F2). In order to analyze the asymptotic behavior of $\mathcal{F}_\mu(x)$ for large $x$, we first note that for any fixed value of $\mu$, $z(x)\to 0$ as $x\to \infty$ since according to assumption (M1), $x^2\sigma'(x)\to \sigma_1 = 2m/r_H$. Next, we fix the constants $z_1$ and $\delta$ in Lemma~\ref{Lem:Technical2}(a) which only depend on the equation of state. For any fixed value of $\mu$ and large enough $x$ we then have
$$
\mathcal{F}_\mu(x) \geq \delta^2 z(x)^{2/3}
\left[ 1 + 4a(x) \right] - 1
 = \delta^2\frac{(2|\mu|)^{2/3}}{(x^2\sigma'(x))^{1/3}}
\left[ \frac{1}{x} + \frac{4\sigma(x)}{x^2\sigma'(x)} \right] - 1\\
 \to 4\delta^2 \left(\frac{2|\mu|}{\sigma_1^2}\right)^{2/3} - 1.
$$
Therefore, we can guarantee that $\mathcal{F}_\mu(x)$ is positive for large $x$ if we choose $|\mu|$ large enough such that $|\mu| > \sigma_1^2/(16 \delta^3)$. In this case, the smooth function $\mathcal{F}_\mu$ is negative for $x=1$ and positive for large $x$, and thus it must have a zero by the intermediate value theorem.

Next, we show that this zero must be unique. For this, we consider the derivative of $\mathcal{F}_\mu$:
\begin{equation}
\mathcal{F}_\mu'(x) = \nu(z(x))^2
\left\{ \frac{w(z(x))}{x a(x)}[1+ 4a(x)][x a'(x) - 1 - 4 a(x)] + 4a'(x) \right\}.
\end{equation}
It follows from our assumptions (M4) and (F3) that the right-hand side is always positive. Indeed, when $w=0$ the expression inside the curly parenthesis reduces to $4a'(x) > 0$, while for the other extreme, $w = 1/3$, it reduces to
\begin{displaymath}
\frac{1}{3 x a} \left[ (1+4a)(xa' - 1 - 4a) + 12x a a' \right]
 = \frac{1}{3 x a} \left[ (1+16a) xa' - (1 + 4a)^2 \right],
\end{displaymath}
which is also positive as a consequence of Lemma~\ref{Lem:Technical1}. Therefore, the function $\mathcal{F}_\mu$ is strictly monotonously increasing, and as a consequence, the critical points of the dynamical system are unique.

After having determined the critical point we analyze the flow in the vicinity of it. To this purpose, we linearize the Hamiltonian vector field $X_F$ at the critical point $(x_c,z_c)$. The linearization is given by the matrix
\begin{equation}
D X_F(x_c,z_c) = \left( 
\begin{array}{cc}
 \frac{\partial^2 F_\mu(x_c,z_c) }{\partial x\partial z} &
 \frac{\partial^2 F_\mu(x_c,z_c) }{\partial z^2} \\  
 -\frac{\partial^2 F_\mu(x_c,z_c) }{\partial x^2} &
 -\frac{\partial^2 F_\mu(x_c,z_c) }{\partial x\partial z}
\end{array} \right)
 = \sigma'(x_c) \frac{f(z_c)^2}{z_c}\left(
\begin{array}{cc}
 2 & \frac{\eta}{\Lambda} \\
 \frac{\Lambda}{a(x_c)}\left[ x_c a'(x_c) - 1 - 4a(x_c) \right] & -2
\end{array} \right),
\label{Eq:DynSysLin}
\end{equation}
where $\Lambda := z_c/x_c$ and $\eta := 1 - \nu(z_c)^2 + w(z_c) > 0$. A short calculation which exploits the fact that $\nu(z_c)^2[1 + 4a(x_c)] = 1$ reveals that the determinant is
\begin{equation}
\det\left[ D X_F(x_c,z_c) \right]
 = -x_c\sigma'(x_c)^2\frac{f(z_c)^4}{z_c^2} \mathcal{F}_\mu'(x_c) < 0.
\end{equation}
Since $DX_F(x_c,z_c)$ has zero trace, it follows that its eigenvalues are real and have different signs; in particular $(x_c,z_c)$ is hyperbolic and the Hartman-Grobman linearization theorem~\cite{Hartman-Book} is applicable. Furthermore, the negative and positive eigenvalues of $D_X F(x_c,z_c)$ give rise to one-dimensional stable and unstable local manifolds, corresponding, respectively, to the set of points $(x,z)$ converging to the critical point when $\lambda\to \infty$ and $\lambda\to -\infty$. Accordingly, the function $F_\mu$ has a saddle point at $(x_c,z_c)$, since the determinant of $D X_F(x_c,z_c)$ is equal to the one of the Hessian of $F_\mu$ at $(x_c,z_c)$. In a vicinity of $(x_c,z_c)$, the critical point together with the stable and unstable manifolds describe the level set of $F_\mu$ through the critical point. In the following, we study the extensions of the stable and unstable manifolds to the whole phase space $\Omega$.

%%%%%%%%%%%%%%%%%%%%%%%%%%%%%%%%%%%%%%%%%%%%
\subsection{Extension of the stable and unstable local manifolds and global properties of the phase space}
\label{SubSec:DSGlobal}

In the previous subsection, we showed that for large enough $|\mu|$ there exists a critical point of the dynamical system~(\ref{Eq:DynSys}), and further we showed that a critical point is always unique and hyperbolic, with corresponding one-dimensional stable and unstable local manifolds $\Gamma_-$ and $\Gamma_+$. In this subsection, we discuss the extensions of these manifolds towards the horizon ($x=1$) and towards the asymptotic region $x\to\infty$, and we discuss the global qualitative behavior of the phase flow.

Our result is based on the analysis of the two curves $\Gamma_1$ and $\Gamma_2$, corresponding to the points $(x,z)$ with vanishing partial derivative of $F_\mu(x,z)$ with respect to $x$ and $z$, respectively. Using Eqs.~(\ref{Eq:CritCond1},\ref{Eq:CritCond2}), these two curves are therefore parametrized by
\begin{equation}
z_1(x) := \frac{2|\mu|}{\sqrt{ x^5\sigma'(x)}},\qquad
z_2(x) := H^{-1}(x^4\sigma(x)),\quad x > 1,
\end{equation}
with the function $H: (0,\infty) \to (0,\infty)$ defined as
\begin{equation}
H(z) := \frac{\mu^2}{z^2}\left(\frac{1}{\nu(z)^2}  -1\right), \quad z > 0.
\label{Eq:Hz}
\end{equation}
Notice that $H$ is invertible since $H(z)\to \infty$ as $z\to 0$ and $H(z)\to 0$ as $z\to \infty$, and since
\begin{displaymath}
\frac{dH(z)}{dz} = -\frac{2\mu^2}{z^3\nu^2(z)}\left[ 1 - \nu^2(z) +w(z) \right] < 0,
\end{displaymath}
which shows that $H(z)$ is monotonously decreasing. Furthermore, using Lemma~\ref{Lem:Technical1},
\begin{eqnarray*}
\frac{dz_1}{dx}(x) &=& \frac{z_1(x)}{2x}\frac{xa'(x) - 4a(x) - 1}{a(x)} < 0,\\
\frac{dz_2}{dx}(x) &=& \frac{x^3}{\frac{dH}{dz}\left( H^{-1}(x^4\sigma(x) \right) }
\left[ x\sigma'(x) + 4\sigma(x) \right] < 0.
\end{eqnarray*}
Therefore, $z_1(x)$ and $z_2(x)$ are monotonously decreasing functions. The curves $\Gamma_1$ and $\Gamma_2$ intersect themselves only at the unique critical point $(x_c,z_c)$, and therefore, they divide the phase space into four regions, see Fig.~\ref{Fig:Fase} below. In each of these four regions, the flow has a particular direction which is indicated by the arrows in this figure. We note in passing that the curve $\Gamma_2$ has a nice physical interpretation: it divides the phase space into regions where the flow is sub- and supersonic, respectively. In order to see this, recall the expression~(\ref{Eq:RadialVelocity}) for the radial velocity of the flow measured by a static observer. In terms of dimensionless quantities,
$$
\frac{|v|}{c} = \frac{|\mu|}{\sqrt{\mu^2 + x^4\sigma'(x) z^2}}.
$$
Comparing this expression with
$$
\nu(z_2(x)) = \frac{|\mu|}{\sqrt{\mu^2 + x^4\sigma'(x) z_2(x)^2}},
$$
which follows from the definition of $z_2(x)$, and using the monotonicity of $\nu(z)$, this implies that $|v|/c < \nu$ for points $(x,z)$ lying above the curve $\Gamma_2$, while $|v|/c > \nu$ for points $(x,z)$ lying below this curve. Since $z_2(x)\to\infty$ as $x\to 1$, an important conclusions is that any differentiable solution $z(x)$ of the implicit equation~(\ref{Eq:FundamentalBis}) which extends from the event horizon to $x\to \infty$ must be transonic. As discussed in Sec.~\ref{Sec:Model} this solution must cross the curve $\Gamma_2$ precisely at the critical point, and hence it must coincide with either the stable or unstable local manifold.

\begin{figure}[htp]
\begin{center}
\includegraphics[width=10.0cm]{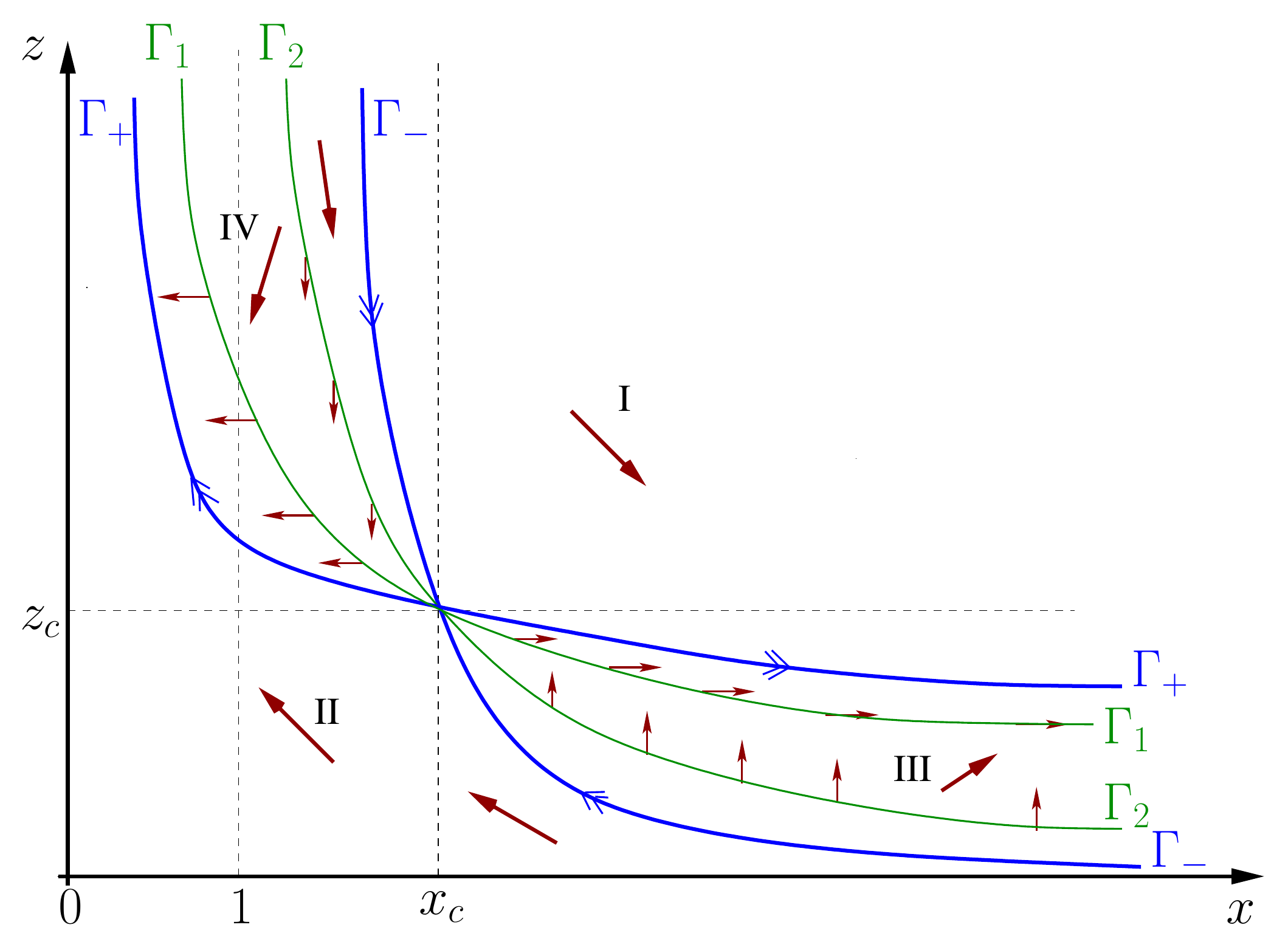}
\end{center}
\caption{\label{Fig:Fase} A sketch of the phase space, showing the critical point $(x_c,z_c)$, the stable ($\Gamma_-$) and unstable ($\Gamma_+$) manifolds, and the curves $\Gamma_1$ and $\Gamma_2$ which delimit the four regions I--IV. Also shown is the direction of the flow in each of the regions and along the curves $\Gamma_1$ and $\Gamma_2$. The flow's radial velocity measured by static observers is subsonic in the region above the curve $\Gamma_2$ and supersonic in the region below $\Gamma_2$.}
\end{figure}

In order to proceed we compare the locations of the stable ($\Gamma_-$) and unstable ($\Gamma_+$) local manifolds with those of the curves $\Gamma_1$ and $\Gamma_2$ in the phase space $\Omega$. For this, we first compute the slopes of these four curves at the critical point. The slopes of $\Gamma_\pm$ at the critical point are determined by the eigenvectors of the linearization $D X_F(x_c,z_c)$. From Eq.~(\ref{Eq:DynSysLin}) one finds the eigenvectors
$$
e_\pm = \left( \begin{array}{c}
 1 \\ -\frac{2\Lambda}{\eta}(1 \mp k)
 \end{array} \right).
$$
corresponding to the eigenvalues $\pm 2\sigma'(x_c)f(z_c)^2 k/z_c$, where
$$
k = \sqrt{1 + \frac{\eta}{4a_c}\left[ x_c a'(x_c) - 1 - 4a(x_c) \right]}.
$$
Since $D X_F(x_c,z_c)$ has negative determinant, $k$ is real and we can take the positive root, and according to Lemma~\ref{Lem:Technical1}, $k < 1$. Therefore, the slopes of the stable and unstable local manifolds are given by
\begin{equation}
l_- = -\frac{2\Lambda}{\eta}(1 + k),\qquad
l_+ = -\frac{2\Lambda}{\eta}(1 - k),
\label{Eq:Brslopes}
\end{equation}
respectively. The slopes of $\Gamma_1$ and $\Gamma_2$ can be obtained by direct computation; the result is
\begin{equation}
l_1 := \frac{dz_1}{dx}(x_c) = -\frac{2\Lambda}{\eta}(1-k)(1+k),
\label{Eq:z1slope}
\end{equation}
and
\begin{equation}
l_2 := \frac{dz_2}{dx}(x_c) = -\frac{2\Lambda}{\eta}.
\label{Eq:z2slope}
\end{equation}
Eqs.~(\ref{Eq:Brslopes},\ref{Eq:z1slope},\ref{Eq:z2slope}) and the fact that $0 < k < 1$ imply the inequalities
\begin{equation}
l_- < l_2 <  l_1 < l_+ < 0.
\end{equation}
Therefore, close to the critical point, the relative location of $\Gamma_1$, $\Gamma_2$, $\Gamma_+$ and $\Gamma_-$ is as shown in Fig.~\ref{Fig:Fase}. In particular, the curves $\Gamma_1$ and $\Gamma_2$ lie between the stable and unstable local manifolds, at least in a vicinity of the critical point. In the following, we show that this is also true globally, that is, $\Gamma_\pm$ do not intersect the curves $\Gamma_1$ and $\Gamma_2$ at points other than the critical one. In order to prove this we make the following general observations:
\begin{itemize}
\item[(I)] Consider region I, which is the region that lies above the curves $\Gamma_1$ and $\Gamma_2$. Here, $\frac{\partial F}{\partial x} > 0$ and $\frac{\partial F}{\partial z} > 0$, implying that $x$ increases and $z$ decreases along the flow of the dynamical system~(\ref{Eq:DynSys}). Moreover, at the boundary points $x > x_c$, $z = z_1(x)$, the flow is pointing \emph{inward}, implying that it cannot leave region I at those points. As a consequence, the unstable manifold $\Gamma_+$ must extend to $x\to \infty$, since according to Eq.~(\ref{Eq:HamVec}) $x$ increases and $z$ decreases but is bounded from below by $z > z_1(x)$.

Using similar arguments for the time-reversed flow, one concludes that the stable manifold $\Gamma_-$ extends from $(x_c,z_c)$ to $z\to\infty$ inside region I, with $x$ decreasing monotonically.
\item[(II)] Next, consider region II which lies below the curves $\Gamma_1$ and $\Gamma_2$. Here, $\frac{\partial F}{\partial x} < 0$ and $\frac{\partial F}{\partial z} < 0$, implying that $x$ decreases and $z$ increases along the flow of the dynamical system~(\ref{Eq:DynSys}). At the boundary points $x < x_c$, $z = z_1(x)$ the flow is pointing \emph{inward}, implying that $\Gamma_+$ cannot leave region II at those points. Since $x$ decreases and $z$ increases, and since the curve $\Gamma_1$ crosses the horizon at $x=1$, it follows that $\Gamma_+$ extends from $(x_c,z_c)$ to a point $(1,z_h)$ where it crosses the horizon, with $z_c < z_h < z_1(1)$.

By a similar argument, one can show that the stable manifold $\Gamma_-$ extends  in the past from $(x_c,z_c)$ to $x\to \infty$, $z\to 0$, since it must lie between the $x$-axis and the curve $\Gamma_2$.
\end{itemize}
In conclusion, we have shown that the \emph{unstable} manifold extends from the the asymptotic region $x\to \infty$ through the critical point to the horizon, where $z(1) = z_h$ has a finite value. The corresponding flow is \emph{subsonic} for $x > x_c$ and \emph{supersonic} for $x < x_c$. In contrast, the \emph{stable} manifold extends from $x\to \infty$ through the critical point, but $z(x)$ diverges as $x\to 1$ approaches the horizon. The corresponding flow is \emph{supersonic} for $x > x_c$ and \emph{subsonic} for $x < x_c$. Other possible flows are given by integral curves which lie below the curves $\Gamma_+$ and $\Gamma_-$, whose corresponding flow is everywhere supersonic, and integral curves which lie above the curves $\Gamma_+$ and $\Gamma_-$, whose flow is everywhere subsonic. In all cases, $z$ increases as $x$ decreases and hence, one can associate to each value of the dimensionless radius $x = r/r_H$ a unique value $n(r) = n_0 z(x)$ of the particle density, which is a monotonously decreasing function of $r$.

In order to conclude the proof of Theorem~\ref{Thm:Main} it remains to analyze the asymptotic behavior of the functions $z_\pm$, parametrizing the manifolds $\Gamma_\pm$, for $x\to\infty$. Let us start with $z_+$. Since $z_+$ is monotonously decreasing and since $z_+(x) > z_1(x)$ for all $x > x_c$ the limit
$$
z_\infty = \lim\limits_{x\to\infty} z_+(x)
$$
exists. As we argue now, this limit cannot be zero. Indeed, $z_+(x) > z_1(x)$ for all $x > x_c$ implies that along $\Gamma_+$ the radial velocity is bounded by
$$
|u| = \frac{|\mu|}{x^2 z_+} < \frac{|\mu|}{x^2 z_1(x)} 
 = \frac{\sqrt{x^2\sigma'(x)}}{2\sqrt{x}}
$$
in the asymptotic region. According to assumption (M1), the numerator converges to a positive constant for $x\to\infty$, implying that $u\to 0$ as $x\to \infty$. Therefore, since $F_\mu$ is constant along $\Gamma_+$,
\begin{equation}
F_\mu(x_c,z_c) = F_\mu(x_c,z_+(x_c)) = \lim\limits_{x\to\infty} F_\mu(x,z_+(x))
 = f(z_\infty)^2,
\label{Eq:Matching}
\end{equation}
where we have used Eq.~(\ref{Eq:Fmu}) in the last step. Lemma~\ref{Lem:LProperties} below shows that $F_\mu(x_c,z_c) > 1$, implying that $z_\infty$ must be positive. Therefore, along $\Gamma_+$ the particle density $n(r) = n_0 z(x)$ converges to a finite, positive value and the radial component $u^r$ of the four-velocity and the fluid's velocity $v$ measured by static observers converge to zero when $r\to \infty$, as expected.

In contrast to this, along the stable manifold $\Gamma_-$ we must have
$$
\lim\limits_{x\to\infty} z_-(x) = 0,
$$
since $0 < z_-(x) < z_2(x)\to 0$ as $x\to \infty$, implying that
$$
1 < F_\mu(x_c,z_c) = \lim\limits_{x\to\infty} F_\mu(x, z_-(x) ) = 1 + u_\infty^2,
$$
and while the particle density $n(r) = n_0 z(x)$ converges to zero, $u^r$ and $v$ converge to non-zero values as $r\to \infty$.  This concludes the proof of Theorem~\ref{Thm:Main}.

%%%%%%%%%%%%%%%%%%%%%%%%%%%%%%%%%%%%%%%%%%%%
\subsection{The matching problem}
\label{SubSec:Matching}

In the previous subsections, we discussed the local and global properties of the flow for a given value of the dimensionless accretion rate $\mu$. In this subsection, we prove Theorem~\ref{Thm:MainBis} which states that for any given positive value $n_\infty > 0$ of the particle density at infinity there exists a unique value of $\mu$ such that the function $F_\mu$ has a critical point whose function $n_+$ parametrizing the unstable manifold satisfies
$$
\lim\limits_{r\to\infty} n_+(r) = n_\infty.
$$

So let $z_\infty > 0$. According to Eq.~(\ref{Eq:Matching}) we need to find $\mu$ and a critical point $(x_c,z_c)$ of $F_\mu$ such that
$$
F_\mu(x_c,z_c) = f(z_c)^2\frac{\sigma(x_c)}{1 - \nu(z_c)^2} = f(z_\infty)^2,
$$
where we have used Eq.~(\ref{Eq:CritCond1}) in the first equality. Instead of $\mu$, we parametrize the critical point by $z_c$ and use the fact that $\nu(z_c)^2[1 + 4a(x_c)] = 1$ (see Eq.~(\ref{Eq:AuxiliaryFct})) in order to express $z_c$ as a function of $x_c$. Therefore, we introduce the auxiliary function $\mathcal{L}: (0,\infty)\to \Real$ defined as
\begin{equation}
\mathcal{L}(z):=f(z)^2\frac{\sigma(x_c(z)) }{1 - \nu(z)^2},\qquad z > 0,
\label{Eq:DefL}
\end{equation}
with the function $x_c: (0,\infty)\to (1,\infty)$ given by
\begin{equation}
x_c(z) := a^{-1}\left( \frac{1}{4}\left[\nu(z)^{-2} - 1 \right] \right),\qquad z > 0.
\label{Eq:xzDef}
\end{equation}
where the inverse of the function $a: (1,\infty)\to (0,\infty)$ exists according to  Lemma~\ref{Lem:Technical1}. Our goal is to show the existence of a unique $z_c > 0$ such that
\begin{equation}
\mathcal {L}(z_c) = F_\mu(x_c,z_c) = f(z_\infty)^2.
\label{Eq:DefLPuntCri}
\end{equation}
According to the result of Lemma~\ref{Lem:Technical2}(a), the right-hand side is strictly larger than one. The result then follows from:

\begin{lemma}
\label{Lem:LProperties}
The function $\mathcal{L}: (0,\infty)\to \Real$ defined in Eq.~(\ref{Eq:DefL}) is strictly monotonously increasing and satisfies
\begin{equation}
\lim\limits_{z\to 0} \mathcal{L}(z) = 1,\qquad
\lim\limits_{z\to\infty} \mathcal{L}(z) = \infty
\end{equation}
In particular, there exists for any $f_\infty > 1$ a unique $z_c > 0$ such that $\mathcal{L}(z_c) = f_\infty^2$.
\end{lemma}

\proof See Appendix~\ref{App:Proofs}.
\qed

As a consequence of Lemma~\ref{Lem:LProperties}, given $z_\infty > 0$, there exists a unique $z_c > 0$ such that $\mathcal{L}(z_c) = f(z_\infty)^2$. Defining $(x_c,z_c) := (x_c(z_c),z_c)$ and $\mu :=  \pm\sqrt{x_c^5\sigma'(x_c)} z_c/2$ it is not difficult to verify that the function $F_\mu$ defined in Eq.~(\ref{Eq:Fmu}) has a critical point at $(x_c,z_c)$. Since $|\mu|$ depends uniquely on $x_c$ and $z_c$, and $x_c$ depends uniquely on $z_c$, $|\mu|$ is unique. According to Theorem~\ref{Thm:Main} there exists a unique differentiable function $z_+ : (1,\infty)\to\Real$ such that $z_+(x_c) = z_c$, $F_\mu(x,z_+(x)) = F_\mu(x_c,z_c)$ for all $x > 1$ and which has regular limits as $x\to 1$ and $x\to \infty$. Since
$$
f\left( \lim\limits_{x\to\infty} z_+(x) \right) 
 = \lim\limits_{x\to\infty} F_\mu(x,z_+(x)) = F_\mu(x_c,z_c) = \mathcal{L}(z_c) 
 = f(z_\infty)^2,
$$
it follows by the monotonicity of $f$ that
$$
\lim\limits_{x\to\infty} z_+(x) = z_\infty,
$$
as desired.

%%%%%%%%%%%%%%%%%%%%%%%%%%%%%%%%%%%%%%%%%%%%
\section{Examples and counter-examples}
\label{Sec:Examples}
%%%%%%%%%%%%%%%%%%%%%%%%%%%%%%%%%%%%%%%%%%%%

In this section, we discuss several particular examples of accretion flows. We start with the study of the two limiting cases in condition (F2), corresponding to perfect fluids with sound speed $v_s = 0$ and $v_s = c$, respectively. In the first case, $h = e_0$ is constant implying that the gas does not have internal energy nor pressure, and the matter in this case is dust. The second case describes a stiff fluid, for which the enthalpy is proportional to the particle density, $h(n) =  k n$ for some positive constant $k$. Although in both cases the function $F$ defined in Eq.~(\ref{Eq:Fundamental}) does not possess any critical point, there are still non-trivial, globally defined solutions which we discuss below. Next, we consider three numerical examples concerning the radial accretion of a polytropic fluid. The first two examples are for a Schwarzschild background and have adiabatic index $\gamma = 4/3$ and $\gamma = 1.99$, respectively. In the first case, one obtains the standard Michel solution while in the second case which violates the condition (F3) there is no global solution. The third example has adiabatic index $\gamma = 5/3$ and describes a flow on a distorted black hole which violates condition (M4). While there is still a global flow, it has features which are different than in the standard case where all the assumptions are satisfied. Finally, we analyze a radial polytropic fluid flow with $1 < \gamma \leq 5/3$ on an arbitrary static, spherically symmetric black hole satisfying the conditions (M1)--(M4) assuming that the sound speed at infinity $v_\infty$ is much less than the speed of light. In this case, the sonic sphere is located far from the event horizon, $r_c \gg r_H$, and the parameters characterizing the flow, like its accretion rate, compression rates etc. can be expanded in powers of $v_\infty/c$.

\subsection{Accretion of dust}

For dust, $h(n) = e_0$ is equal to the rest energy of the particles, and in this case Eqs.~(\ref{Eq:Fundamental}) and (\ref{Eq:RadialVelocity}) yield the simple relation
\begin{equation}
\frac{\sigma(r)}{1 - \frac{v(r)^2}{c^2}} 
 = \left(\frac{j_\varepsilon}{e_0 j_n}\right)^2 = const.
\label{Eq:FundamentalDust}
\end{equation}
Taking the limit $r\to \infty$ and requiring that $v(r)\to 0$ and $\sigma(r)\to 1$ in this limit, the constant must be equal to one and it follows that the energy and particle fluxes are related to each other by $j_\varepsilon = e_0 j_n$. Then, Eq.~(\ref{Eq:FundamentalDust}) yields the following expression for the fluid's velocity measured by static observers:
\begin{equation}
v(r) = -c\sqrt{1 - \sigma(r)},
\end{equation}
or $v(r) = -\sqrt{2GM/r}$ for a Schwarzschild black hole of mass $M$, which is just the Newtonian expression for a freely falling radial particle with zero velocity at infinity. The particle density can be obtained using Eq.~(\ref{Eq:RadialVelocity}),
\begin{equation}
n(r) = \frac{|j_n|}{4\pi c}\frac{1}{r^2\sqrt{1 - \sigma(r)}},
\label{Eq:nDust}
\end{equation}
or $n(r) = |j_n|/(4\pi \sqrt{2GM r^3})$ in the Schwarzschild case. The condition (M1) for the existence of a positive mass at infinity implies that $n(r)\to 0$ as $r\to\infty$, so the particle density at infinity is zero. In fact, there are no solutions of Eq.~(\ref{Eq:FundamentalDust}) with $n_\infty > 0$: the assumption $n_\infty > 0$ leads to $j_\varepsilon/(e_0 j_n) = 1$ when taking the limit $r\to \infty$ in Eq.~(\ref{Eq:Fundamental}), and it follows that $n(r)$ is again of the form Eq.~(\ref{Eq:nDust}), leading to a contradiction. Therefore, the existence of a radial steady-state fluid accretion flow into a nonrotating black hole requires nonzero pressure for the particle density at infinity to be positive.

We also see from Eq.~(\ref{Eq:nDust}) that there is no global regular solution if $\sigma$ is not restricted to be everywhere strictly smaller than one.

\subsection{Accretion of a stiff fluid}

In the stiff case, $h(n) = k n$ for some positive constant $k$ and Eq.~(\ref{Eq:Fundamental}) yields
\begin{equation}
n^2\sigma(r) + \frac{\mu^2}{r^4} = \left( \frac{j_\varepsilon}{k j_n} \right)^2 = const,
\label{Eq:FundamentalStiff}
\end{equation}
where $\mu = j_n/(4\pi c)$. Taking the limit $r\to\infty$ and assuming again that $\sigma(r)\to 1$ in this limit gives $n_\infty = j_\varepsilon/(k j_n)$, which fixes the relation between the energy and particle fluxes. Next, taking the limit $r\to r_H$ in Eq.~(\ref{Eq:FundamentalStiff}) yields $\mu = -r_H^2 n_\infty$, which determines the particle and energy fluxes to be
\begin{equation}
j_n = -4\pi c r_H^2 n_\infty,\qquad
j_\varepsilon = -4\pi c k(r_H n_\infty)^2.
\end{equation}
With these constants Eq.~(\ref{Eq:FundamentalStiff}) yields the expression
\begin{equation}
n(r) = n_\infty\sqrt{ \frac{1 - \frac{r_H^4}{r^4}}{\sigma(r)} }
\label{Eq:ParticleDensityStiff}
\end{equation}
for the particle density. Using l'H\^opital's rule, we obtain the compression rate
\begin{equation}
\frac{n(r_H)}{n_\infty} = \frac{2}{\sqrt{r_H\sigma'(r_H)}}
\end{equation}
between the particle densities at the horizon and at infinity, which is finite if the condition (M2) is imposed. For the Schwarzschild case,
\begin{equation}
n(r) = n_\infty\sqrt{ \left( 1 + \frac{2m}{r} \right)\left( 1 + \frac{4m^2}{r^2} \right)},
\end{equation}
and the compression rate is $2$. However, it could be considerably lower or higher for black holes exhibiting a different behavior close to the event horizon.

Finally, the velocity measured by static observers is obtained from Eqs.~(\ref{Eq:RadialVelocity}) and (\ref{Eq:ParticleDensityStiff}):
\begin{equation}
v(r) = -c\frac{r_H^2}{r^2},
\end{equation}
which decays as $r^{-2}$ (as opposed to $r^{-1/2}$ in the dust case).

\subsection{Polytropic fluids}

\begin{figure}[htp]
\begin{center}
 \includegraphics[width=8.0cm]{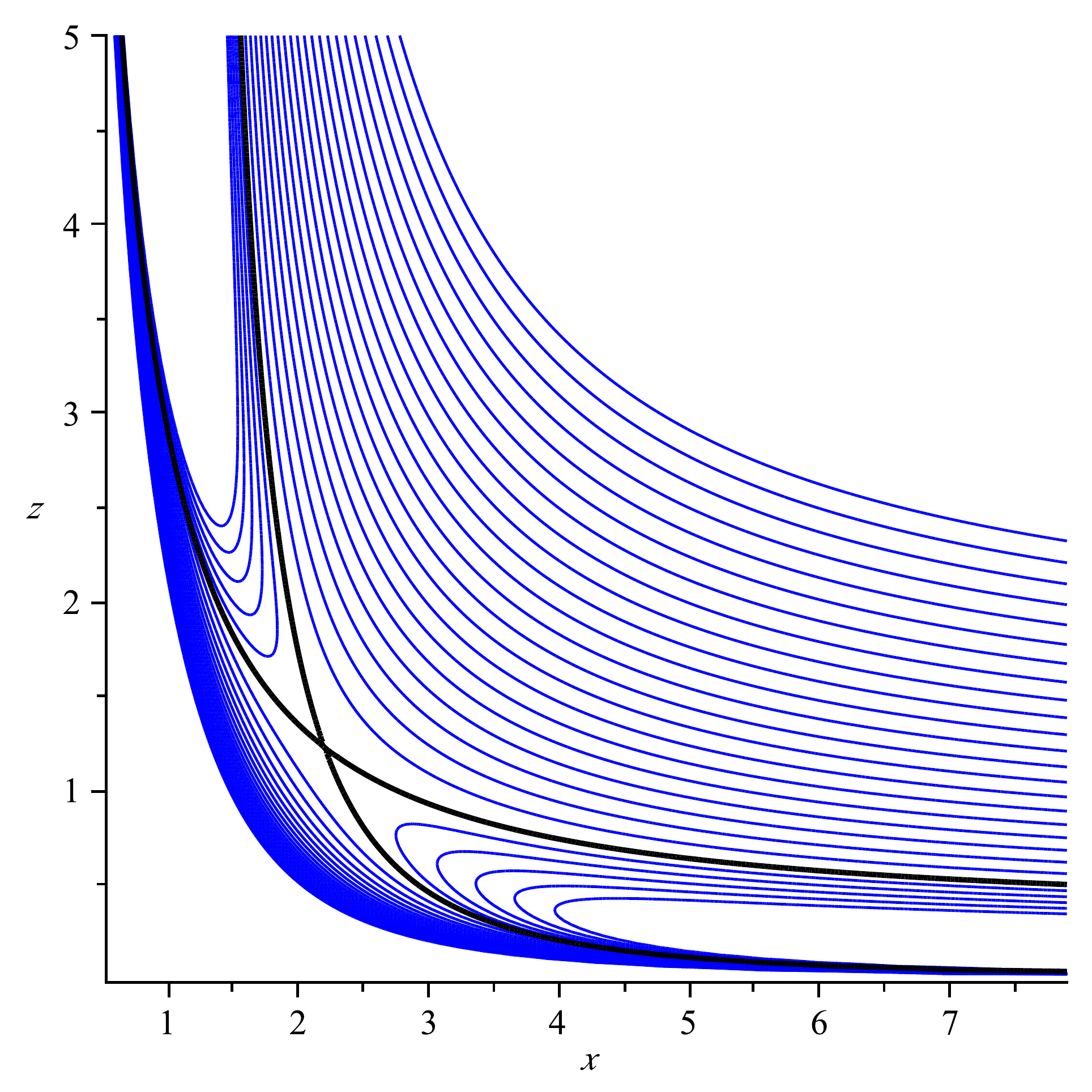}
 \includegraphics[width=8.0cm]{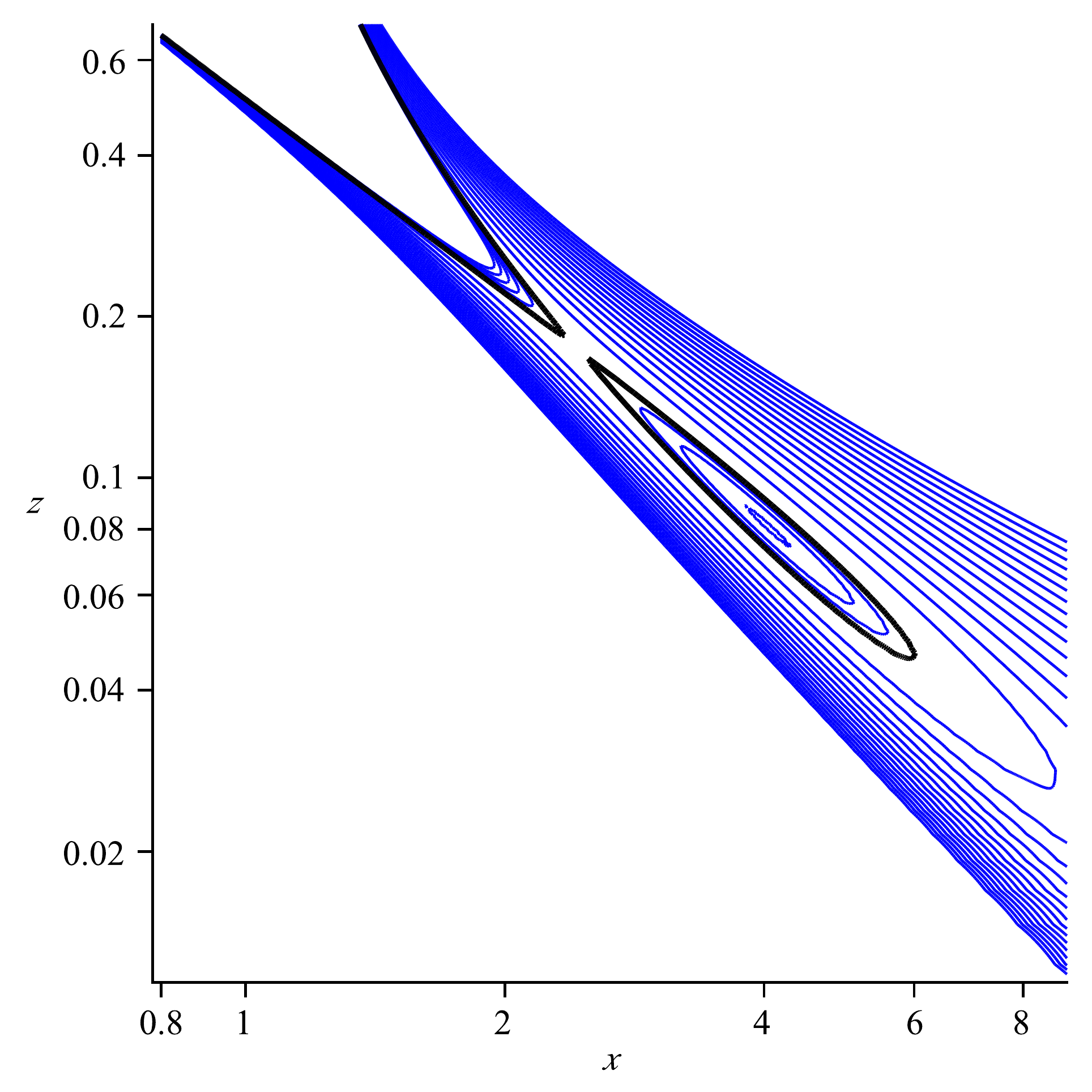}
\end{center}
\caption{\label{Fig:PS} The level sets of the function $F_\mu$ for a polytropic fluid accreted by a Schwarzschild black hole. The level set corresponding to the critical point is displayed with the thick black line. Left panel: In this case, $\gamma = 4/3$ and there is a unique critical point located at $(x_c,z_c) = (2.20,1.23)$ corresponding to a parameter value of $\mu = -2$. The thick black curve is the union of the level curves $(r,n_+(r))$ and $(r,n_-(r))$ described in Theorem~\ref{Thm:Main}, and the former curve which extends from the horizon ($x=1$) to the asymptotic region ($x\to +\infty$) describes the Michel flow. Right panel: In this case, $\gamma = 1.99$ and there are two critical points for $\mu = -0.33$, one corresponds to a saddle point of $F_\mu$ and is located at $(x_c,z_c) = (2.50,0.17)$, while the other one is a local extremum of $F_\mu$ which is located at $(x,z) = (3.7,0.09)$. In this case, the two level curves through the saddle point do not extend to infinity. Instead, they seem to wind around the local extremum and connect to each other. There is no globally-defined radial flow solution in this case.}
\end{figure}

Next, we consider a polytropic equation of state, for which
\begin{equation}
h(n) = e_0 + \frac{\gamma k}{\gamma-1} n^{\gamma-1}
\end{equation}
with $k > 0$ a positive constant, $e_0$ the rest energy of the particle and $\gamma > 1$ the adiabatic index. Condition (F1) is satisfied as long as $e_0 > 0$. The sound speed is
\begin{equation}
v_s(n) = c\sqrt{\frac{\partial\log h}{\partial\log n}} 
 = c\sqrt{\gamma-1}
 \sqrt{ \frac{n^{\gamma-1}}{n^{\gamma-1} + \frac{(\gamma-1)e_0}{\gamma k}}},
\end{equation}
which satisfies condition (F2) as long as $1 < \gamma \leq 2$. The logarithmic derivative of the sound speed is
\begin{equation}
W(n) = \frac{\partial\log v_s}{\partial\log n} 
= \frac{\gamma-1}{2}
\frac{1}{1 + \frac{\gamma k}{(\gamma-1)e_0} n^{\gamma-1}},
\end{equation}
which satisfies the required bound in (F3), provided that $1 < \gamma \leq 5/3$. Defining
\begin{equation}
n_0 := \left[ \frac{(\gamma-1)e_0}{\gamma k} \right]^{1/(\gamma-1)}
\label{Eq:n0DefPoly}
\end{equation}
we obtain, in terms of dimensionless quantities, $z = n/n_0$, and
\begin{eqnarray}
f(z) &=& \frac{h(n)}{e_0} = 1 + z^{\gamma-1},
\label{Eq:fPoly}\\
\nu(z) &=& \frac{v_s(n)}{c} = \sqrt{\gamma-1}\sqrt{\frac{z^{\gamma-1}}{1+z^{\gamma-1}}},
\label{Eq:nuPoly}\\
w(z) &=& \frac{\gamma-1}{2(1 + z^{\gamma-1})}.
\label{Eq:wPoly}
\end{eqnarray}
We provide three examples showing the level sets of the function $F_\mu$ defined in Eq.~(\ref{Eq:Fmu}). The first two examples are shown in Fig.~\ref{Fig:PS} and refer to a Schwarzschild black hole background. In the first example, $\mu = -2$ and $\gamma = 4/3$, and there is a unique hyperbolic critical point which is located at  $(x_c,z_c) = (2.20,1.23)$ through which two level curves of $F_\mu$ cross each other. Only one of these two curves extends from the horizon to infinity, and describes the physical accretion flow. In the second example, $\mu = -0.33$ and $\gamma=1.99$, which violates the upper bound in condition (F3). Here, there are two critical points, one corresponding to a saddle point and the other to a local extremum of $F_\mu$. The right panel of Fig.~\ref{Fig:PS} suggests that the level curves of $F_\mu$ through the saddle point connect to each other instead of extending to the asymptotic region $x\to \infty$. Consequently, there is no global flow in this case. Recently, such ``homoclinic-type" accretion flow solutions have also been found in a cosmological context, see Refs.~\cite{pMeMjK13,pM15}.

\begin{figure}[htp]
\begin{center}
\includegraphics[width=12.0cm]{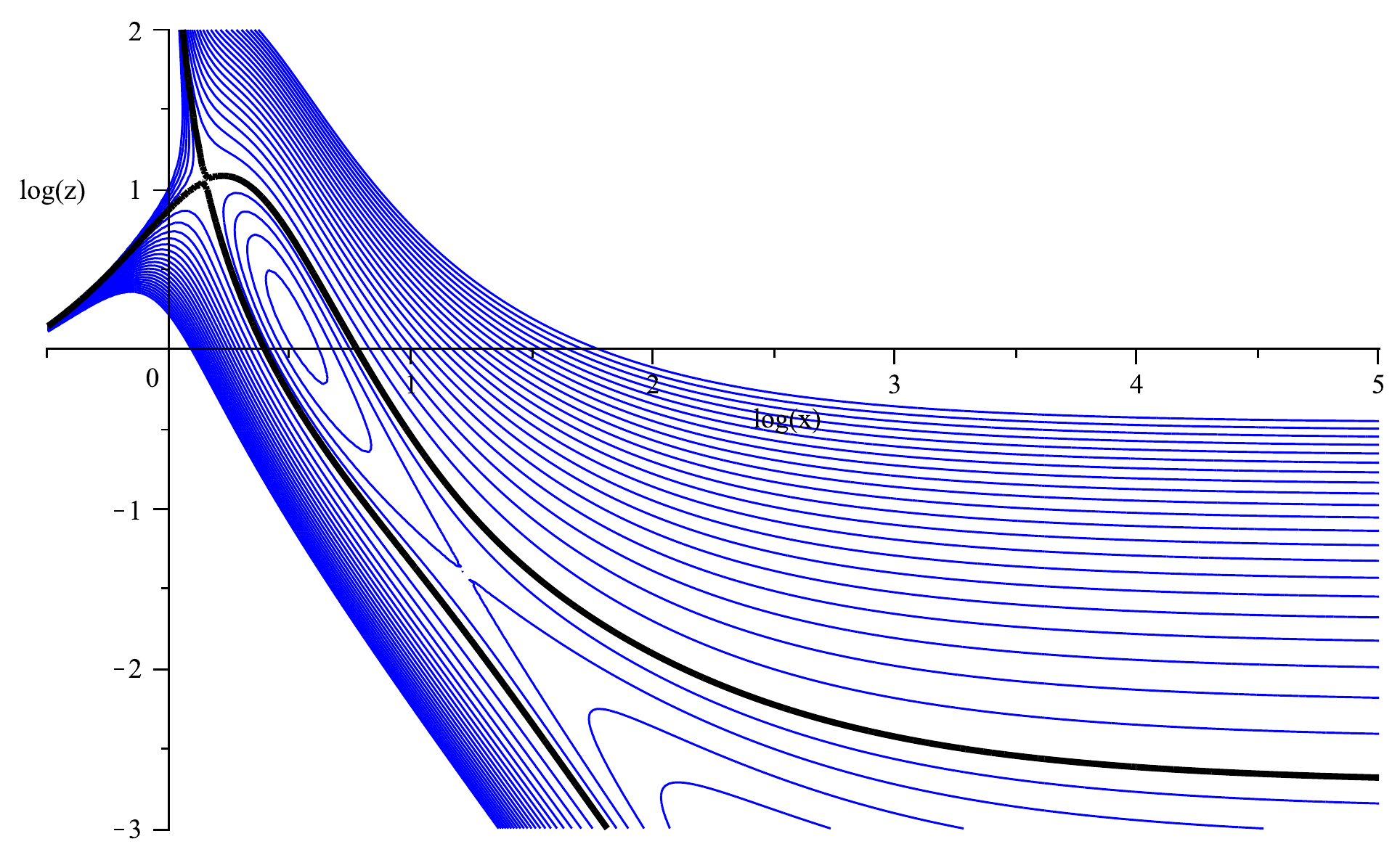}
\end{center}
\caption{\label{Fig:PSBis} The level sets of the function $F_\mu$ for a polytropic fluid with adiabatic index $\gamma = 5/3$ and $\mu = -1$ accreted by a strongly distorted black hole described by Eq.~(\ref{Eq:sigmaDistorted}). In this case, there are three critical points, two of which are located at $(x,z) = (1.16,2.88)$ and $(x,z) = (3.42,0.24)$, respectively, and correspond to a saddle point, and one located at $(x,z) = (1.66,1.19)$ which corresponds to a local extremum of $F_\mu$. The level set corresponding to the saddle point closest to the horizon is displayed with the thick black line. The branch that extends from the horizon to infinity describes a radial, transonic accretion flow. However, as can be seen from the graph, unlike the standard case where the condition (M4) holds, in this case the particle density is not monotonously decaying in $r$, but reaches a maximum value just before the flow passes through the critical point and becomes supersonic.}
\end{figure}

The third example, shown in Fig.~\ref{Fig:PSBis}, has parameter values $\mu = -1$ and $\gamma = 5/3$ and has a background metric described by a distorted black hole, such that
\begin{equation}
\sigma(x) = 1 - \frac{1}{x} - \frac{4}{x^2} + \frac{8}{x^3} - \frac{4}{x^4},\qquad
x \geq 1.
\label{Eq:sigmaDistorted}
\end{equation}
In this case, the conditions (M1)--(M3) on the metric are satisfied, however (M4) is violated. An analysis of the function $\mathcal{F}_\mu(x)$ defined in Eq.~(\ref{Eq:AuxiliaryFct}) reveals the existence of three critical points, two corresponding to a saddle point of $F_\mu$ and one to a local extremum. Unlike the previous example with $\gamma=2$ there does exists a global accretion flow; however, due to the presence of the extremum, the features of this flow are different from the standard situation where all assumptions (M1)-(M4) are satisfies. For example, the particle density ceases to be a monotonously decreasing function of $r$.


\subsection{Small sound speed expansion}

In this subsection, we consider the situation where the sonic sphere lies in the region where the gravitational field is weak, that is, $r_c \gg r_H$. Furthermore, we assume that in the asymptotic region, the sound speed $v_\infty$ is much smaller than the speed of light, such that $\nu_\infty := v_\infty/c \ll 1$. Finally, we also assume that the fluid is described by a polytropic equation of state, as in the previous subsection. With these assumptions, we can expand the metric function
\begin{equation}
\sigma(x) = 1 - \frac{\sigma_1}{x} - \frac{\sigma_2}{x^2}
 + {\cal O}\left( \frac{1}{x^3} \right),\qquad \sigma_1 = \frac{2m}{r_H},
\label{Eq:sigmaExpansion}
\end{equation}
for $x \geq x_c$, and also we can expand the fluid quantities in terms of the small parameter $\nu_\infty$. Our first goal is to express the quantities $x_c$, $z_c$, $\nu_c$, and $\mu$ characterizing the flow in terms of $\nu_\infty$.

\begin{figure}[htp]
\begin{center}
\includegraphics[width=8.0cm]{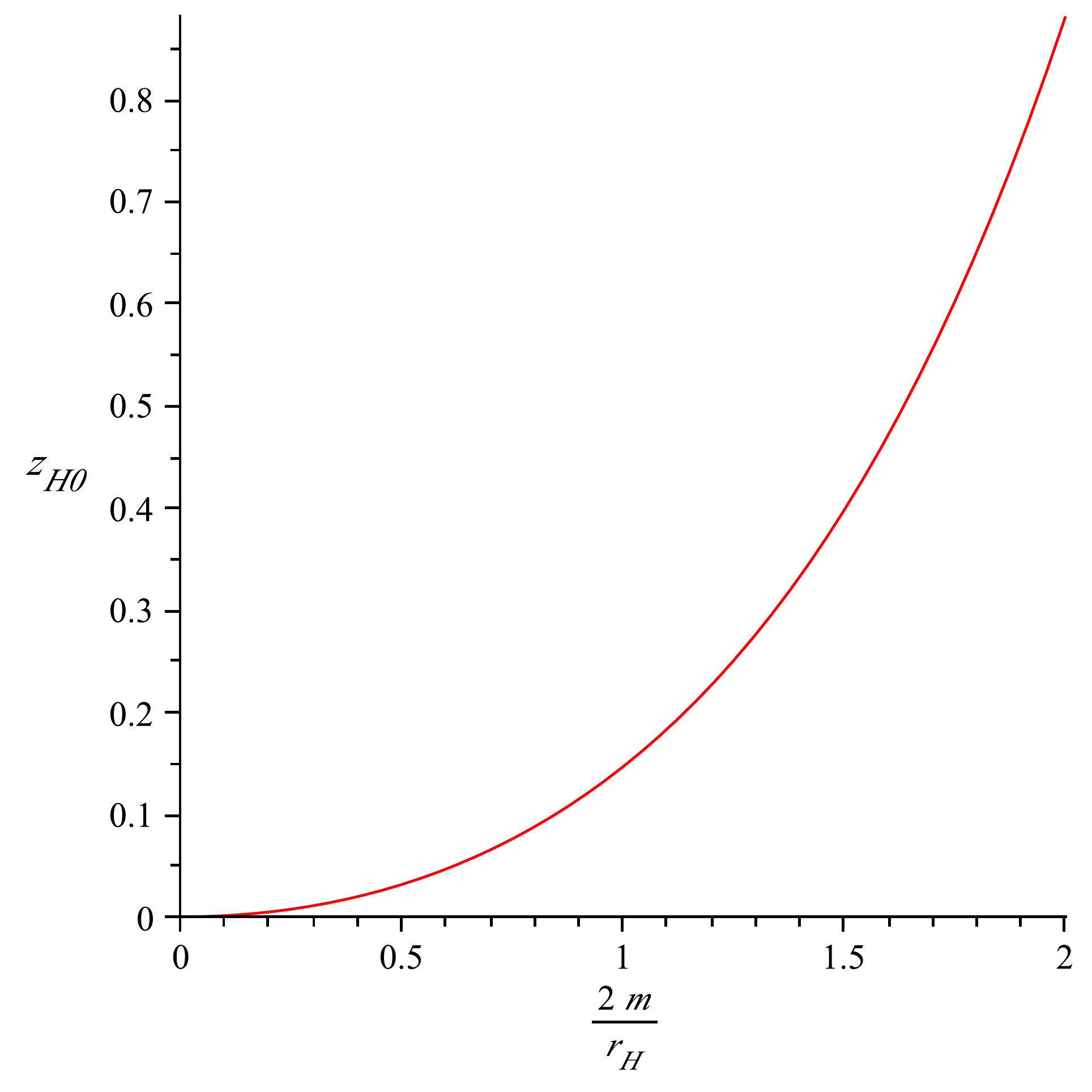}
\end{center}
\caption{\label{Fig:zH0} The parameter $z_{H0}$ vs. $\sigma_1 = 2m/r_H$.}
\end{figure}

We begin with the relation between the particle density at the critical point $z_c$ and the sound speed $\nu_\infty$ at infinity. This relation is determined by the equations~(\ref{Eq:DefL},\ref{Eq:DefLPuntCri}), that is,
\begin{equation}
\mathcal{L}(z_c) = f(z_c)^2\frac{\sigma(x_c(z_c))}{1 - \nu(z_c)^2} = f(z_\infty)^2,
\label{Eq:zczinfty}
\end{equation}
with the functions $f$ and $\nu$ given by Eqs.~(\ref{Eq:fPoly}) and~(\ref{Eq:nuPoly}), and the function $x_c$ defined in Eq.~(\ref{Eq:xzDef}). Using the expansion in Eq.~(\ref{Eq:sigmaExpansion}) we find the following expansion for $x_c(z_c)$:
\begin{equation}
\frac{\sigma_1}{x_c(z_c)} = 4(\gamma-1) z_c^{\gamma-1}\left\{
 1 - \left[ 3\gamma - 2 + 8(\gamma-1)\frac{\sigma_2}{\sigma_1^2}
  \right] z_c^{\gamma-1} 
 + {\cal O}\left( z_c^{2(\gamma-1)} \right) \right\}.
\label{Eq:xcExpansion}
\end{equation}
Substituting this result in Eq.~(\ref{Eq:zczinfty}) we find
\begin{equation}
\mathcal{L}(z_c) = 1 + (5-3\gamma)z_c^{\gamma -1}
 + \left[1+ 3(\gamma-1)(3\gamma-4)+16(\gamma-1)^2\frac{\sigma_2}{\sigma_1^2}\right] z_c^{2(\gamma -1)} + {\cal O}\left( z_c^{3(\gamma -1)} \right).
\label{Eq:DefLExpansion}
\end{equation}
On the other hand, from Eq.~(\ref{Eq:nuPoly}) we obtain
\begin{equation}
z_\infty = \left( \frac{\nu_\infty^2}{\gamma-1} \right)^{\frac{1}{\gamma-1}}
\left[ 1 + \frac{\nu_\infty^2}{(\gamma-1)^2} + {\cal O}(\nu_\infty^4) \right]
\label{Eq:zinfty}
\end{equation}
for $\nu_\infty \ll 1$, and hence
\begin{equation}
f(z_\infty)^2 = (1+z_\infty^{\gamma -1})^2
 = \left(1 + \frac{\nu_\infty^2}{\gamma -1 -\nu_\infty^2}\right)^2
 = 1+ 2\frac{\nu_\infty^2}{\gamma -1} + 3\frac{\nu_\infty^4}{(\gamma -1)^2} 
 + {\cal O}(\nu_\infty^6).
\label{Eq:fExpansion}
\end{equation}
Comparing Eq.~(\ref{Eq:DefLExpansion}) with Eq.~(\ref{Eq:fExpansion}) we conclude that for $1<\gamma\leq 5/3$ the solution $z_c$ of the equation $\mathcal{L}(z_c) = f(z_\infty)^2$, which is unique by the monotonicity of $\mathcal{L}$, must satisfy $z_c\ll 1$. Depending on whether $1<\gamma < 5/3$ or $\gamma = 5/3$ we obtain the following behavior for the quantities $z_c$, $x_c$, $\nu_c$, $z_H$, $z_\infty$  and $\mu$ in terms of $\nu_\infty\ll 1$:
\begin{itemize}
\item Case I ($1 < \gamma < 5/3$):
\begin{eqnarray}
z_c &=& 
\left[\frac{2\nu_\infty^2}{(5-3\gamma)(\gamma -1)}\right]^{\frac{1}{\gamma -1}}
\left[ 1 + A\nu_\infty^2 + {\cal O}(\nu_\infty^4) \right],
\\
x_c &=& \frac{5-3\gamma}{8}\frac{\sigma_1}{\nu_\infty^2}
\left[1 + B\nu_\infty^2 + {\cal O}(\nu_\infty^4) \right],
\\
\nu_c &=& \sqrt{\frac{2}{5-3\gamma}}\nu_\infty
\left[1 + C\nu_\infty^2 + {\cal O}(\nu_\infty^4) \right],
\\
|\mu| &=& \frac{\sigma_1^2}{4}\lambda
\left( \frac{\nu_\infty^{5-3\gamma}}{\gamma-1} \right)^{\frac{1}{\gamma-1}}
\left[1 + D\nu_\infty^2 + {\cal O}(\nu_\infty^4) \right], 
\\
z_H &=& \frac{\sigma_1^2}{4}\lambda
\left( \frac{\nu_\infty^{5-3\gamma}}{\gamma-1} \right)^{\frac{1}{\gamma-1}}\left[1+  \left(D-\frac{1}{\gamma-1}\right)\nu_\infty^2+{\cal O}( \nu_\infty^4) \right]
\end{eqnarray}
with
\begin{displaymath}
\lambda = 2^{\frac{9-7\gamma}{2(\gamma-1)}}
(5-3\gamma)^{-\frac{5-3\gamma}{2(\gamma-1)}}
\end{displaymath}
and where the coefficients $A$, $B$, $C$ and $D$ are defined by
\begin{eqnarray}
A &:=& \frac{1}{(5-3\gamma)^2(\gamma -1)^2}
\left[ 12 - \frac{1}{2}(3\gamma+1)^2 - 32(\gamma-1)^2\frac{\sigma_2}{\sigma_1^2}
\right],
\\
B &:=& \frac{2}{(5-3\gamma)(\gamma-1)}
\left[ 3\gamma-2+8(\gamma-1)\frac{\sigma_2}{\sigma_1^2}\right]
 -(\gamma-1)A
 =  \frac{1}{2}\frac{1}{(5-3\gamma)^2} 
 \left[ 9(7-3\gamma) + 32(3-\gamma)\frac{\sigma_2}{\sigma_1^2} \right],
\\
C &:=& \frac{1}{2}(\gamma-1)A-\frac{1}{(5-3\gamma)(\gamma-1)}
 = -\frac{1}{4}\frac{1}{(5-3\gamma)^2}
 \left[ 3(3\gamma+1) + 64(\gamma-1)\frac{\sigma_2}{\sigma_1^2} \right],
 \\
 D &:=& A + \frac{3}{2} B + \frac{8}{5-3\gamma}\frac{\sigma_2}{\sigma_1^2}
 = \frac{1}{4}\frac{1}{(5-3\gamma)(\gamma -1)^2} 
 \left[ 27\gamma^2 - 66\gamma + 47 + 64(\gamma-1)^2\frac{\sigma_2}{\sigma_1^2} \right].
\end{eqnarray}
Here, $x_c := x_c(z_c)$ and $\nu_c := \nu(z_c)$ have been computed by substituting the expansion for $z_c$ into Eq.~(\ref{Eq:xcExpansion}) and Eq.~(\ref{Eq:nuPoly}), respectively. The dimensionless accretion rate is obtained by substituting the expansions for $x_c$ and $z_c$ into the expression $2|\mu| = \sqrt{x_c^5\sigma'(x_c)} z_c$. The density at the horizon $z_H$ is computed from the relation between $z_H$ and $z_\infty$ which follows from
$$
(1+z_H^{\gamma-1})^2\frac{\mu^2}{z_H^2} = F_\mu(1,z_H) = f(z_\infty)^2
 = (1 + z_\infty^{\gamma-1})^2,
$$
leading to the equation
\begin{equation}
\frac{|\mu|}{z_H} (1+z_H^{\gamma-1}) = 1 + z_\infty^{\gamma-1}
 = 1 + \frac{\nu_\infty^2}{\gamma -1} + \frac{\nu_\infty^4}{(\gamma -1)^2} 
 + {\cal O}(\nu_\infty^6).
\label{Eq:zH}
\end{equation}
Since the dimensionless accretion rate $|\mu|$ converges to zero as fast as $\nu_\infty^{(5-3\gamma)/(\gamma-1)}$ for $\nu_\infty\to 0$, $z_H$ must converge to zero at the same rate. With this observation in mind, the expansion of $z_H$ in terms of $\nu_\infty$ follows easily from Eq.~(\ref{Eq:zH}).
\item Case II ($\gamma = 5/3$):
\begin{eqnarray}
z_c &=& \left(\frac{\nu_\infty}{l}\right)^{3/2}
\left[ 1 + {\cal O}(\nu_\infty) \right],
\\
x_c &=& \frac{3}{8}\sigma_1\frac{l}{\nu_\infty}
\left[ 1 + {\cal O}(\nu_\infty) \right],
\\
\nu_c &=&  \sqrt{\frac{2}{3}} \sqrt{ \frac{\nu_\infty}{l} }
\left[ 1 + {\cal O}(\nu_\infty) \right],
\\
|\mu| &=& \frac{\sigma_1^2}{2}
\left(\frac{3}{8} \right)^{3/2}
\left[ 1 + {\cal O}(\nu_\infty) \right],
\\
z_{H}&=& z_{H0}[1+{\cal O}(\nu_\infty)], 
\end{eqnarray}
where
\begin{equation}
l := \sqrt{1 + \left( \frac{4}{3} \right)^3\frac{\sigma_2}{\sigma_1^2}},
\label{Eq:l}
\end{equation}
and $z_{H0}$ is uniquely determined by the equation
\begin{equation}
z_{H0} - 3\beta(z_{H0}^{2/3} + 1) = 0,\qquad
\beta := \frac{\sigma_1^2}{6}\left( \frac{3}{8} \right)^{3/2}
\label{Eq:zH0}
\end{equation}
whose solution can be written explicitly as
\begin{equation}
z_{H0} = \left( \beta + w + \frac{\beta^2}{w} \right)^3,\qquad
w = \beta^{1/3}\left( \beta^2 + \frac{3}{2} + \sqrt{3}\sqrt{\beta^2 + \frac{3}{4}} \right)^{1/3}.
\label{Eq:zH0Solution}
\end{equation}
A plot of $z_{H0}$ as a function of $\sigma_1$ is shown in Fig.~\ref{Fig:zH0}. In contrast to case I, here the accretion rate converges to a finite, positive value as $\nu_\infty\to 0$ and consequently, it follows from Eq.~(\ref{Eq:zH}) that $z_H$ must also converge to a finite positive value $z_{H0}$ when $\nu_\infty\to 0$. This value is determined by Eq.~(\ref{Eq:zH0}) which follows from Eq.~(\ref{Eq:zH}) by taking the limit $\nu_\infty\to 0$. In the Schwarzschild case, $\ell=1$ and $z_{H0} \simeq 0.14677$.
\end{itemize}

We see that in both cases the critical point is located far from the event horizon, $r_c \gg r_H$ when $\nu_\infty\ll 1$, which is consistent with our assumption of the critical point lying in the weak field region. Finally, we compute the accretion rate $j_\varepsilon$ and the compression rates $n_c/n_\infty$ and $n_H/n_\infty$ the gas undergoes when it falls from infinity through the critical point into the black hole. These quantities can be obtained by combining the expansions above for $\mu$, $z_c$ and $z_H$ with Eq.~(\ref{Eq:zinfty}). The result is
\begin{itemize}
\item Case I ($1 < \gamma < 5/3$):
\begin{eqnarray}
j_\varepsilon &=& 4\lambda\pi c e_0\left( \frac{GM}{c^2} \right)^2
n_\infty\nu_\infty^{-3}
\left[ 1 + \left( D + \frac{\gamma-2}{(\gamma-1)^2} \right)\nu_\infty^2
 + {\cal O}(\nu_\infty^4) \right],\\
\frac{z_c}{z_\infty}&=&\left( \frac{2}{5-3\gamma}\right)^{\frac{1}{\gamma -1}}
\left[1 + \left(A -\frac{1}{(\gamma-1)^2} \right)\nu_\infty^2+ {\cal O}(\nu_\infty^4) \right],
\\
\frac{z_H}{z_\infty}&=&\lambda\frac{\sigma_1^2}{4} \nu_\infty^{-3}\left[1+\left(D+\frac{\gamma-2}{(\gamma-1)^2}\right)\nu_\infty^2+ {\cal O}(\nu_\infty^4) \right].
\end{eqnarray}
\item Case II ($\gamma = 5/3$):
\begin{eqnarray}
j_\varepsilon &=& \pi c e_0\left( \frac{GM}{c^2} \right)^2 n_\infty
\nu_\infty^{-3}[1+{\cal O}(\nu_\infty)],\\
\frac{z_c}{z_\infty}&=&\left( \frac{2}{3l}\right)^{3/2}\nu_\infty^{-3/2}[1+{\cal O}(\nu_\infty)],
\\
\frac{z_H}{z_\infty}&=&\left(\frac{2}{3}\right)^{3/2}z_{H0}\nu_\infty^{-3}[1+{\cal O}(\nu_\infty)],
\end{eqnarray}
\end{itemize}
where we have used $r_H\sigma_1 = 2GM/c^2$ with $M$ the mass of the black hole. We see that in both cases the accretion rate scales like $M^2 n_\infty\nu_\infty^{-3}$ and that $z_H$ is larger than $z_\infty$ by a factor which is proportional to $\nu_\infty^{-3}$, implying that significant compression rates are achieved during the accretion process. On the other hand, in case I the compression rate at the sonic sphere is almost independent of $\nu_\infty$ for small $\nu_\infty$, while it scales like $\nu_\infty^{-3/2}$ in case II. In both cases the compression rate at the horizon becomes larger as $\sigma_1 = 2m/r_H$ increases.

%%%%%%%%%%%%%%%%%%%%%%%%%%%%%%%%%%%%%%%%%%%%
\section{Conclusions}
\label{Sec:Conclusions}
%%%%%%%%%%%%%%%%%%%%%%%%%%%%%%%%%%%%%%%%%%%%

In this work, we have studied the Michel flow, describing the steady spherically symmetric accretion of a perfect fluid into a nonrotating black hole, under the assumption that the fluid's density is sufficiently small so that its self-gravity can be neglected. Our treatment does not assume a polytropic equation of state nor a Schwarzschild background geometry. Instead, we imposed rather general conditions on the fluid equation of state and considered a large class of static, spherically symmetric and asymptotically flat black holes with a regular horizon. Consequently, our results are also applicable for certain deformed Schwarzschild black holes which might arise in alternative theories of gravity or in classical general relativity in the presence of external matter fields such as dark matter, for example.

Our main result is that for a given equation of state and metric satisfying our assumptions, there exists for each positive $n_\infty > 0$ a unique, spherical and stationary solution of the general relativistic Euler equations which is everywhere regular at and outside the event horizon and whose particle density at infinity is $n_\infty$. This solution describes a transonic flow, the flow being supersonic close to the black hole and subsonic in the asymptotic region. We have also computed the accretion rate and the compression rates the fluid undergoes at the event horizon and the sonic sphere, assuming a polytropic equation of state and the usual assumption that the fluid's speed of sound $v_\infty$ is much smaller than the speed of light at infinity. While to leading order in $v_\infty/c$ the accretion rate is given by the familiar Bondi formula and only depends on the background metric through its total mass, the compression rate at the horizon depends on the details of the metric fields describing the black hole; in particular it depends on the ratio $r_s/r_H$ between its Schwarzschild radius $r_s = 2m$ and the radius of the event horizon $r_H$. For a Schwarzschild black hole, $r_s = r_H$ and this ratio is one. However, a spherically symmetric static black hole in the presence of external fields satisfying the weak energy condition has $r_s > r_H$ (see Eq.~(\ref{Eq:Gtt})), leading to higher compression rates at the horizon than in the absence of external fields. Intuitively, this might be expected since small black holes have large surface gravity, and in this sense the gravitational acceleration at the horizon is larger when $r_H < r_s$ than in the Schwarzschild case, resulting in higher compression rates.

Further, we have given several examples for the accretion flow problem, using different background metrics and equations of state, and analyzed cases where our assumptions are not satisfied. We have shown that violations of the assumptions on the fluid equation of state may lead to situations where the presence alone of a critical point is not sufficient to guarantee the existence of a globally defined accretion flow, while strongly distorted black holes violating our assumptions on the metric may lead to accretion flows which have distinct features compared to the standard case where all the assumptions are satisfied.

There are various interesting questions that remain to be studied. For example, it would be desirable to consider more realistic fluids, where effects due to viscosity and emission of radiation are taken into account. Moreover, it should also be interesting to consider fluctuations away from the precise spherical and steady-state assumptions made in this work. Spherical and nonspherical linear perturbations of the Michel flow have been analyzed by Moncrief~\cite{vM80} and shown to remain bounded outside the sonic sphere, thus establishing the linear stability of the Michel flow in the subsonic region. It would be interesting to extend this stability result to the region between the black hole horizon and the sonic sphere. In Ref.~\cite{eCmMoS15} we analyze small spherical and nonspherical acoustic perturbations of the Michel flow and show that these perturbations exhibit quasi-normal oscillations.

%%%%%%%%%%%%%%%%%%%%%%%%%%%
%%%   ACKNOWLEDGMENTS
%%%%%%%%%%%%%%%%%%%%%%%%%%%

\acknowledgments

It is our pleasure to thank Adriana Gazol, Luis Lehner, Patryk Mach, N\'estor Ortiz, and Thomas Zannias for fruitful and stimulating discussions. We thank the gravitational physics group at University of Vienna, where part of this work was performed, for their hospitality. OS also thanks the Perimeter Institute for Theoretical Physics for hospitality. This research was supported in part by CONACyT Grants No. 238758, 101353, 233137, by a CIC Grant to Universidad Michoacana and by Perimeter Institute for Theoretical Physics. Research at Perimeter Institute is supported by the Government of Canada through Industry Canada and by the Province of Ontario through the Ministry of Research and Innovation.

%%%%%%%%%%%%%%%%%%%%%%%%%%%%%%%%%%%%%%%%%%%%
\appendix
\section{Justification for the metric conditions (M1)--(M4)}
\label{App:MetricConditions}
%%%%%%%%%%%%%%%%%%%%%%%%%%%%%%%%%%%%%%%%%%%%

In this appendix we provide some additional details regarding the physical motivation for the metric conditions (M1)--(M4) given in Sec.~\ref{Sec:Model}. For definitions and discussions of Komar mass, Killing horizons, surface gravity and the energy conditions used below we refer the reader to Refs.~\cite{Heusler-Book,Wald-Book}.

Condition (M1) implies that the metric is asymptotically flat and that its Komar mass is well-defined and positive. The Komar mass is defined as
$$
M_{Komar} = \lim\limits_{r\to \infty} -\frac{c^2}{8\pi G} \int_{s_r}\star d\underline{\bf k}=\lim\limits_{r\to \infty}\frac{c^2}{2G}\frac{r^2\sigma'(r)}{\alpha(r)},
$$
where $S_r$ denotes a sphere of radius $r$, $\underline{\bf k} = {\bf g}({\bf k},\cdot) = g_{\mu\nu} k^\mu dx^\nu = -\sigma c dt$ is the one-form associated with the Killing vector ${\bf k}$, and $\star$ denotes the Hodge dual. According to our assumptions in (M1) the limit exists and $M_{Komar} = c^2 m/G$. Next, condition (M2) implies that the Killing vector field ${\bf k}$ is null at the surface ${\cal H} := \{ r = r_H \}$. Since $\alpha(r_H) > 0$ we also have $N(r_H) = 0$ which shows that the surface ${\cal H}$ is null. Therefore, ${\cal H}$ is a Killing horizon, and its associated surface gravity $\kappa$, defined by $\left. d\sigma \right|_{\cal H}= -2c^{-2}\kappa \left. \underline{\bf k} \right|_{\cal H}$ is given by
$$
\kappa = \frac{c^2}{2}\frac{\sigma'(r_H)}{\alpha(r_H)},
$$
which is strictly positive according to our assumption.

Next, we provide justification for condition (M3), assuming that the stress-energy tensor $T_{\mu\nu}^{(0)}$ associated with the metric in Eq.~(\ref{Eq:metric}) via Einstein's equations satisfies suitable energy conditions. The $tt$, $rr$ and $\vartheta\vartheta$ components of Einstein's equations yield, respectively,
\begin{eqnarray}
m' &=& \frac{4\pi G}{c^4} r^2 \varepsilon^{(0)},
\label{Eq:Gtt}\\
\frac{N}{r}\frac{\sigma'}{\sigma} &=& \frac{2m}{r^3} + \frac{8\pi G}{c^4} p_r^{(0)},
\label{Eq:Grr}\\
\frac{1}{2\alpha r^2}\frac{d}{dr}\left( \frac{r^2}{\alpha}\sigma' \right)
 - \frac{N}{r}\frac{\alpha'}{\alpha} &=& \frac{8\pi G}{c^4} p_t^{(0)},
\label{Eq:GAA}
\end{eqnarray}
where $m(r)$ is the Misner-Sharp mass function, defined by $N = 1 - 2m/r$, and $\varepsilon^{(0)}$, $p_r^{(0)}$ and $p_t^{(0)}$ refer to the energy density, radial and tangential pressure of the effective stress-energy tensor $T_{\mu\nu}^{(0)}$. (The superscript $^{(0)}$ indicates that these quantities refer to the effective stress-energy tensor $T_{\mu\nu}^{(0)}$ defined from the background metric as opposed to the stress-energy tensor $T_{\mu\nu}$ of the accretion flow.) Combining Eqs.~(\ref{Eq:Gtt},\ref{Eq:Grr},\ref{Eq:GAA}) we also obtain the following equations:
\begin{eqnarray}
\frac{N}{r}\frac{\alpha'}{\alpha} &=& \frac{4\pi G}{c^4}(\varepsilon^{(0)} + p_r^{(0)}).
\label{Eq:alpha}\\
\frac{1}{\alpha r^2}\frac{d}{dr}\left( \frac{r^2}{\alpha}\sigma' \right) 
 &=& \frac{8\pi G}{c^4}\left( \varepsilon^{(0)} + p_r^{(0)} + 2p_t^{(0)} \right),
\label{Eq:dsigma}
\end{eqnarray}
The \emph{weak energy condition}, stating the all physical observers should measure a nonnegative energy density, implies that $\varepsilon^{(0)}\geq 0$ and $\varepsilon^{(0)} + p_r^{(0)}\geq 0$. Therefore, under this condition, Eq.~(\ref{Eq:Gtt}) and the fact that $m(r_H) = r_H/2 > 0$ imply that the mass function $m(r)$ is positive for all $r\geq r_H$, and Eq.~(\ref{Eq:alpha}) and $\alpha\to 1$ for $r\to\infty$ yield $\alpha(r) > 0$ for all $r\geq r_H$. Under the additional assumption that the radial pressure is nonnegative, Eq.~(\ref{Eq:Grr}) also implies that the function $\sigma$ is strictly monotonically increasing, justifying condition (M3). Instead of assuming $p_r^{(0)} \geq 0$, it is also possible to resort to the \emph{strong energy condition}, which states that $\varepsilon^{(0)} + p_r^{(0)} + 2p_t^{(0)}\geq 0$. Eq.~(\ref{Eq:dsigma}) then implies that the function $J(r) := r^2\sigma'(r)/\alpha(r)$ is monotonically non-decreasing, which, together with $J(r_H) = 2r_H^2\kappa/c^2 > 0$ implies that $\sigma'(r) > 0$ for all $r\geq r_H$.

Finally, we make some comments regarding condition (M4). First, we remark that the strong and weak energy conditions and Eqs.~(\ref{Eq:alpha},\ref{Eq:dsigma}) imply that $(r^2\sigma')' \geq 0$, such that the first inequality in Eq.~(\ref{Eq:M4}) is automatically satisfied. Next, consider the simple example in which $\sigma$ is a second-order polynomial in $1/r$,
$$
\sigma(r) = 1 - \frac{2m}{r} + \frac{e m^2}{r^2},
$$
with $e < 1$ a parameter. The horizon is located at $r_H = m[ 1 + \sqrt{1 - e}]$ and
$$
r^2\sigma'(r) = 2m - \frac{2em^2}{r} \geq 0
$$
for all $r\geq r_H$. The condition (M4) is equivalent to
$$
2e - 3\frac{r}{m} <  0 < \frac{9}{8 + \frac{m}{r}}\left( 1 - e\frac{m}{r} \right) - e,\qquad r > r_H, 
$$
which is satisfied for all negative values of $e$ and all small enough positive values such that
$$
e < \frac{9\sqrt{1-e}}{9 + 8\sqrt{1-e}}(1 + \sqrt{1 - e}).
$$

%%%%%%%%%%%%%%%%%%%%%%%%%%%%%%%%%%%%%%%%%%%%
\section{Proofs of Lemmata \ref{Lem:Technical1}, \ref{Lem:Technical2} and \ref{Lem:LProperties} }
\label{App:Proofs}
%%%%%%%%%%%%%%%%%%%%%%%%%%%%%%%%%%%%%%%%%%%%

This appendix is devoted to the proofs of the technical results described in Lemmata~\ref{Lem:Technical1}, \ref{Lem:Technical2} and \ref{Lem:LProperties}.

\subsection{Proof of Lemma~\ref{Lem:Technical1}}

In order to prove Lemma~\ref{Lem:Technical1}, we first note that the conditions (M1)--(M3) imply that $a$ is a well-defined, nonnegative function satisfying $a(1) = 0$ and $\lim_{x\to\infty} a(x)/x = \lim_{x\to\infty} \sigma(x)/(x^2\sigma'(x)) = r_H/2m$. Next, a short computation reveals that
$$
x^2\frac{d}{dx}\left( \frac{a}{x} \right) = 1 - \frac{(x^2\sigma')'}{x\sigma'} a.
$$
Using condition (M4) we conclude from this that
\begin{equation}
1 + 3a > x^2\frac{d}{dx}\left( \frac{a}{x} \right) 
 > 1 - \frac{9\overline{m}}{8 + \frac{\overline{m}}{x}}\frac{a}{x},
\label{Eq:aIneq}
\end{equation}
where we have set $\overline{m} := m/r_H$. The first inequality is equivalent to $xa' < 1 + 4a$. We claim that, as a consequence of the second inequality, $a(x) < x/(2\overline{m})$ for all $x > 1$. If not, there must exist a point $x^* > 1$ for which $1/(2\overline{m}) \leq a(x^*)/x^* \leq 2/(3\overline{m})$ and $d(a(x^*)/x^*)/dx\leq 0$ since $a(x)/x\to 1/(2\overline{m})$ as $x\to\infty$. However, it follows from the second inequality in Eq.~(\ref{Eq:aIneq}) that
$$
\left. x^2\frac{d}{dx}\left( \frac{a}{x} \right) \right|_{x=x^*}
 > 1 - \frac{9\overline{m}}{8}\frac{a(x^*)}{x^*} \geq \frac{1}{4} > 0,
$$
leading to a contradiction. Therefore, $a(x) < x/(2\overline{m})$ for all $x > 1$. Using this result and again the second inequality in Eq.~(\ref{Eq:aIneq}) we find
$$
xa' > 1 + \frac{ 8a - 8a\frac{\overline{m}}{x}}{8 + \frac{\overline{m}}{x}}
 > 1 + \frac{8a - 8a\frac{1}{2a}}{8 + \frac{1}{2a}} = \frac{(4a+1)^2}{16a + 1} > 0,
$$
and the lemma is proved.

\subsection{Proof of Lemma~\ref{Lem:Technical2}}

\begin{enumerate}
\item[(a)] By condition (F3), $w = \frac{\partial \log\nu(z)}{\partial \log z}\leq 1/3$. Integrating both sides from $z$ to $z_1$ with $0 < z\leq z_1 $ we obtain
$$
\log\left(\frac{\nu(z_1)}{\nu(z)}\right) \leq \frac{1}{3}\log\left(\frac{z_1}{z}\right)
 = \log\left(\frac{z_1}{z}\right)^{1/3},
$$
from which the first inequality follows with $\delta := \nu(z_1)/z_1^{1/3}$. For the second inequality, we use this result and the definition of the dimensionless sound speed $\nu(z)$:
$$
\frac{\partial\log f(z)}{\partial\log z} = \nu(z)^2 \geq \delta^2 z^{2/3},
\qquad 0 < z\leq z_1.
$$
Integrating both sides from $z_0$ to $z$ with $0 < z_0 < z \leq z_1$ yields
$$
\log \left(\frac{f(z)}{f(z_0)}\right) \geq 
 \left. \frac{3}{2}\delta^2 z^{2/3} \right|_{z_0}^z.
$$
Taking the limit $z_0\to 0$ and observing that $f(z_0)\to 1$ by (F1), the second inequality follows.
\item[(b)] From (a) we have $f(z)\geq e^{\frac{3}{2}\nu(z_1)^2 (z/z_1)^{2/3}}$ for all $0 < z < z_1$. Taking the limit $z_1\to z$ yields  $f(z) \geq e^{\frac{3}{2}\nu(z)^2} \geq 1$ for all $z > 0$. Since $f(z)\to 1$ as $z\to 0$, it follows that
$$
\lim\limits_{z\to 0} \nu(z) = 0,
$$
as claimed. Next, it follows from (F3) that $\nu$ is a monotonously increasing function which is bounded from above by one according to condition (F2). Therefore, when $z\to \infty$, $\nu$ converges to a finite value which is smaller than or equal to one.
\item[(c)] By the definition of $\nu$ and its monotonicity, we have for all $z\geq z_0 > 0$,
$$
\frac{\partial \log f}{\partial \log z} = \nu^2(z) \geq \nu^2(z_0).
$$
Integrating, we obtain from this
$$
f(z) \geq f(z_0)\left(\frac{z}{z_0}\right)^q,\qquad z > z_0,
$$
where $q := \nu(z_0)^2$, which completes the proof of the lemma.
\end{enumerate}

\subsection{Proof of Lemma~\ref{Lem:LProperties}}

When $z\to 0$, $f(z)\to 1$ and $\nu(z)\to 0$ according to assumption (F1) and Lemma~\ref{Lem:Technical2}(b), implying in particular that $x_c(z)\to \infty$. As a consequence of condition (M1) we obtain $\sigma(x_c(z))\to 1$, and ${\cal L}(z) \to 1$. On the other hand, when $z\to \infty$ we have $f(z)\to \infty$ and $\nu(z)\to \nu_1\leq 1$ according to Lemma~\ref{Lem:Technical2}(c) and (b). If $\nu_1 < 1$, $x_c(z)$ converges to a value larger than one, implying that $\sigma(x_c(z))$ converges to a value larger than zero. Consequently, $\mathcal{L}(z)\to \infty$ as $z\to \infty$. If $\nu_1 = 1$ we have $x_c(z)\to 1$ and $\sigma(x_c(z))\to 0$. However, in this case we can use L'H\^opital's rule and the relation
\begin{equation}
z\frac{dx_c}{dz}(z) = -\frac{1}{2a'(x_c(z))}\frac{w(z)}{\nu(z)^2}
\label{Eq:dxc}
\end{equation}
to conclude that
$$
\lim\limits_{z\to\infty} \frac{\sigma(x_c(z))}{1-\nu(z_c)^2} 
 = \lim\limits_{z\to\infty} \frac{\sigma'(x_c(z))}{4a'(x_c(z))}\frac{1}{\nu(z)^4}
 = \frac{\sigma'(1)}{4} > 0.
$$
Together with $f(z)\to\infty$ this implies again that $\mathcal{L}(z)\to \infty$.

In order to prove the monotonicity statement we compute the derivative of $\mathcal{L}(z)$. Using the definitions of $\nu$ and $w$ and the relations~(\ref{Eq:dxc}) and
$$
\frac{x_c(z)\sigma'(x_c(z)}{\sigma(x_c(z))} = \frac{1}{a(x_c(z))} = \frac{4\nu(z)^2}{1 - \nu(z)^2},
$$
we obtain
\begin{equation}
\frac{d\mathcal{L}}{dz}(z) = 2\sigma(x_c(z))\frac{f(z)^2}{z}
 \left( \frac{\nu(z)}{1-\nu(z)^2} \right)^2
 \left[ 1 - \nu(z)^2 + w(z) - \frac{1}{x_c(z) a'(x_c(z))}\frac{w(z)}{\nu(z)^2} \right],\quad z > 0.
\end{equation}
Using Lemma~\ref{Lem:Technical1} and $4a(x_c(z)) = \nu(z)^{-2} -1$ we find
$$
w(z) - \frac{1}{x_c(z) a'(x_c(z))}\frac{w(z)}{\nu(z)^2} >
w(z) - \frac{16a(x_c(z)) + 1}{[4a(x_c(z))+1]^2}\frac{w(z)}{\nu(z)^2}
 = -3(1 - \nu(z)^2)w(z).
$$
Since $\nu(z)^2\leq 1$ and $0 \leq w\leq 1/3$ it follows that the derivative of $\mathcal{L}(z)$ is strictly positive, implying its strict monotonicity.

\bibliographystyle{unsrt}
\bibliography{../References/refs_accretion}

\begin{thebibliography}{10}

\bibitem{Shapiro-Book}
S.L. Shapiro and S.A. Teukolsky.
\newblock {\em Black Holes, White Dwarfs, and Neutron Stars}.
\newblock John Wiley \& Sons, New York, 1983.

\bibitem{EHT}
Event horizon telescope, http://www.eventhorizontelescope.org.

\bibitem{sDetal08}
S.~Doeleman and et~al.
\newblock Event-horizon-scale structure in the supermassive black hole
  candidate at the galactic centre.
\newblock {\em Nature}, 455:78, 2008.

\bibitem{aBtJaLdP14}
A.E. Broderick, T.~Johannsen, A.~Loeb, and D.~Psaltis.
\newblock Testing the no-hair theorem with event horizon telescope observations
  of {Sagittarius A$^*$}.
\newblock {\em Astrophys. J.}, 784:7, 2014.

\bibitem{hB52}
H.~Bondi.
\newblock On spherically symmetrical accretion.
\newblock {\em Monthly Notices Roy Astronom. Soc.}, 112:195--204, 1952.

\bibitem{fM72}
F.C. Michel.
\newblock Accretion of matter by condensed objects.
\newblock {\em Astrophysics and Space Science}, 15:153--160, 1972.

\bibitem{vM80}
V.~Moncrief.
\newblock Stability of stationary, spherical accretion onto a {S}chwarzschild
  black hole.
\newblock {\em Astrophys. J.}, 235:1038--1046, 1980.

\bibitem{dAsBtD14a}
D.B. Ananda, S.~Bhattacharya, and T.K. Das.
\newblock Acoustic geometry through perturbation of mass accretion rate {I} -
  radial flow in general static spacetime.
\newblock 2014.
\newblock arXiv:1406.4262.

\bibitem{jKeM13}
J.~Karkowski and E.~Malec.
\newblock Bondi accretion onto cosmological black holes.
\newblock {\em Phys.Rev. D}, 87:044007, 2013.

\bibitem{pMeM13}
P.~Mach and E.~Malec.
\newblock Stability of relativistic {B}ondi accretion in
  {S}chwarzschild-(anti-)de {S}itter spacetimes.
\newblock {\em Phys.Rev. D}, 88:084055, 2013.

\bibitem{pMeMjK13}
P.~Mach, E.~Malec, and J.~Karkowski.
\newblock Spherical steady accretion flows: {D}ependence on the cosmological
  constant, exact isothermal solutions, and applications to cosmology.
\newblock {\em Phys.Rev. D}, 88:084056, 2013.

\bibitem{fGfL11}
F.S. Guzm\'an and F.D. Lora-Clavijo.
\newblock Exploring the effects of pressure on the radial accretion of dark
  matter by a {S}chwarzschild supermassive black hole.
\newblock {\em Mon. Not. R. Astron. Soc.}, 415:225--234, 2011.

\bibitem{eM99}
E.~Malec.
\newblock Fluid accretion onto a spherical black hole: Relativistic description
  versus {B}ondi model.
\newblock {\em Phys. Rev. D}, 60:104043, 1999.

\bibitem{fLmGfG14}
F.D. Lora-Clavijo, M.~Gracia-Linares, and F.S. Guzm\'an.
\newblock Horizon growth of supermassive black hole seeds fed with collisional
  dark matter.
\newblock {\em Mon. Not. Roy. Astron. Soc.}, 443:2242--2251, 2014.

\bibitem{vF09}
V.~Faraoni.
\newblock The {L}agrangian description of perfect fluids and modified gravity
  with an extra force.
\newblock {\em Phys. Rev. D}, 80:124040, 2009.

\bibitem{eP65}
E.N. Parker.
\newblock Dynamical properties of stellar coronas and stellar winds. {IV}.
  {T}he separate existence of subsonic and supersonic solutions.
\newblock {\em Astrophys. J.}, 141:1463--1478, 1965.

\bibitem{eCoS12}
E.~Chaverra and O.~Sarbach.
\newblock Polytropic spherical accretion flows on {S}chwarzschild black holes.
\newblock {\em AIP Conf.Proc.}, 1473:54--58, 2012.

\bibitem{Eliana-Master-thesis}
E.~Chaverra.
\newblock {\em Accretion of Matter in Spherical Symmetry (Master Thesis)}.
\newblock UMSNH, Mexico, 2011.

\bibitem{Hartman-Book}
P.~Hartman.
\newblock {\em Ordinary Differential Equations (2nd ed.)}.
\newblock SIAM Classics in Applied Mathematics 38, Philadelphia, 2002.

\bibitem{Perko-Book}
L.~Perko.
\newblock {\em Differential Equations and Dynamical Systems (3rd ed.)}.
\newblock Springer-Verlag, New York, 2001.

\bibitem{pM15}
P.~Mach.
\newblock Homoclinic accretion solutions in the {S}chwarzschild-anti-de
  {S}itter spacetime.
\newblock {\em Phys. Rev. D}, 91:084016, 2015.

\bibitem{eCmMoS15}
E.~Chaverra, M.D. Morales, and O.~Sarbach.
\newblock Quasi-normal acoustic oscillations in the {M}ichel flow.
\newblock {\em Phys. Rev. D}, 91:104012, 2015.

\bibitem{Heusler-Book}
M.~Heusler.
\newblock {\em Black Hole Uniqueness Theorems}.
\newblock Cambridge University Press, Cambridge, England, 1996.

\bibitem{Wald-Book}
R.M. Wald.
\newblock {\em General Relativity}.
\newblock The University of Chicago Press, Chicago, London, 1984.

\end{thebibliography}
%%%%%%%%%%%%%%%%%%%%%%%%%%%%%%%%%%%%%%%%%%%%

\end{document}